\pdfoutput=1
\documentclass[journal,a4paper,10pt]{IEEEtran}
\usepackage{amsmath}
\usepackage{amsfonts}
\usepackage{amssymb}
\usepackage{cite}
\usepackage{multicol}
\usepackage{verbatim}
\usepackage{graphicx}
\usepackage{balance}
\usepackage{epstopdf}
\usepackage{epsfig,color}
\usepackage{subcaption}
\usepackage[linesnumbered,ruled]{algorithm2e}

\SetKwInOut{Parameter}{Parameters}
\SetKwRepeat{Do}{do}{while}

\IEEEoverridecommandlockouts

\begin{document}

\title{Impersonation Detection in Line-of-Sight Underwater Acoustic Sensor Networks}

\author{ Waqas Aman\authorrefmark{1}, Muhammad Mahboob Ur Rahman\authorrefmark{1}, Junaid Qadir\authorrefmark{1}, Haris Pervaiz\authorrefmark{2}, Qiang Ni\authorrefmark{3} \\
\authorblockA{
\authorrefmark{1}Department of Electrical engineering, Information Technology University, Lahore, Pakistan} \\
\authorblockA{
\authorrefmark{2}Institute of Communication Systems, Home of 5GIC, University of Surrey, Guildford, UK} \\
\authorblockA{
\authorrefmark{3}School of Computing and Communications, Lancaster University, Lancaster, UK \\
Corresponding author: Waqas Aman (waqas.aman@itu.edu.pk) }
}

\maketitle 


\begin{abstract}

This work considers a line-of-sight underwater acoustic sensor network (UWASN) consisting of $M$ underwater sensor nodes randomly deployed according to uniform distribution within a vertical half-disc (the so-called trusted zone). The sensor nodes report their sensed data to a sink node on water surface on a shared underwater acoustic (UWA) reporting channel in a time-division multiple-access (TDMA) fashion, while an active-yet-invisible adversary (so-called Eve) is present in the close vicinity who aims to inject malicious data into the system by impersonating some Alice node. To this end, this work first considers an additive white Gaussian noise (AWGN) UWA channel, and proposes a novel, multiple-features based, two-step method at the sink node to thwart the potential impersonation attack by Eve. Specifically, the sink node exploits the noisy estimates of the distance, the angle of arrival, and the location of the transmit node as device fingerprints to carry out a number of binary hypothesis tests (for impersonation detection) as well as a number of maximum likelihood hypothesis tests (for transmitter identification when no impersonation is detected). We provide closed-form expressions for the error probabilities (i.e., the performance) of most of the hypothesis tests. We then consider the case of a UWA with colored noise and frequency-dependent pathloss, and derive a maximum-likelihood (ML) distance estimator as well as the corresponding Cramer-Rao bound (CRB). We then invoke the proposed two-step, impersonation detection framework by utilizing distance as the sole feature. Finally, we provide detailed simulation results for both AWGN UWA channel and the UWA channel with colored noise. Simulation results verify that the proposed scheme is indeed effective for a UWA channel with colored noise and frequency-dependent pathloss.

\end{abstract}

\section{Introduction}

Underwater acoustic sensor networks (UWASN) are utilized by a multitude of civilian and military applications, e.g., sensing a specific area for resources, intrusion detection for border surveillance, and exploration of life underwater \cite{Akyildiz:AdNetworks:2005},\cite{Felemban:IJDSN:2015}. In contrast to the terrestrial wireless networks, the UWASNs are exposed to the peculiar challenges of the underwater acoustic (UWA) channel, e.g. frequency-selective nature of path-loss and ambient noise, severe multipath (longer delay spreads), battery constraints, low (and variable) propagation speed of acoustic waves, and low data rates (for long-range communication) \cite{Akyildiz:AdNetworks:2005}, \cite{Chen:Networks:2018}. The aforementioned challenges make the UWA channel quite error-prone, which calls for design of intelligent forward error correction (FEC) schemes, and retransmission schemes (e.g. ARQ) \cite{Chen:Networks:2018} tailored for UWASNs.

The broadcast nature of the UWA channel also makes the UWASNs vulnerable to various kinds of security breaches by nearby malicious nodes. Traditionally, the broadcast channels (e.g., terrestrial wireless, underwater acoustic) were secured via cryptography-based solutions at higher layers, where mutual trust is established a priori by pre-distributing a set of shared secret keys among the network entities. Recently there has been tremendous interest in complementing the crypto-based security mechanisms at the higher layers with the feature-based security mechanisms at the physical layer \cite{Wang:ICST:2017}. Physical-layer security schemes build upon the so-called features (derived from the propagation medium's characteristics, or, hardware imperfections) to exploit them as \textit{virtual keys} to enforce an additional layer of security in the network \cite{Wang:ICST:2017},\cite{Xianbin:CommMag:2016}.

Various kinds of attacks by adversaries have been investigated in the literature---e.g., impersonation (or, intrusion) attacks, eavesdropping attacks, Sybil attacks, denial-of-service attacks, wormhole attacks, jamming attacks, man-in-the-middle attacks, and malicious relaying---and a detailed survey of these attacks can be seen in the recent survey articles \cite{Wang:ICST:2017},\cite{Han:CommMag:2015},\cite{Domingo:WiComm:2011}. Most importantly, each physical-layer security scheme, like its higher layer counterpart, could counter only certain attacks (and not all of them) while making certain a priori assumptions about Eve (e.g., how much computational and infrastructural resources are at the disposal of Eve), which if violated by Eve renders the scheme ineffective \cite{Wang:ICST:2017},\cite{Xianbin:CommMag:2016}.

This work considers a UWASN whereby a set of $M$ sensor nodes reports its sensed data to a sink node (on the water surface) in a time-division multiple access (TDMA) fashion, while a malicious node Eve is present in the close vicinity. This work assumes an active Eve. When Eve actively transmits, it may either announce its presence by executing a jamming attack, or it may remain in stealth mode to execute an impersonation attack. This work assumes that Eve remains in stealth mode only. That is, Eve---being a clever impersonator and not a mere jammer---wants to deceive the sink node by assuring it that Eve is indeed a legitimate sensor node. This way, Eve could potentially inject malicious data into the system to corrupt the system's data integrity.       

\vspace{2mm}
{\bf Contributions.} The main contributions of this work are:

(C1) This work presents a novel, multiple-features based, two-step method for impersonation detection in an additive white Gaussian noise (AWGN)-limited, line-of-sight UWA channel. The first step implements a binary hypothesis test to enforce a proximity-based authentication. To this end, the sink node exploits the distance estimate of the sender node to determine whether the transmit node lies within a trusted zone (a half-disc of radius $d_0$) or not. The second step assumes that the sink node has the estimate of angle of arrival (AoA), and thus, the estimate of sender node's position available. The estimates of distance, AoA, and position are then exploited as fingerprints of the transmit device, and each of them is passed on to a maximum likelihood test followed by a binary hypothesis test. The individual binary decisions---\textit{impersonation} or \textit{no impersonation}---of all the tests in the second step are fused together (and the fusion outcome is further fused with the binary decision from the first step) to generate the ultimate binary decision. 

(C2) As a by-product, the proposed method also performs transmitter identification when no impersonation is detected in the system.

(C3) Next, we relax the two main assumptions in (C1) that the UWA channel is AWGN, and the distance estimate is available to the sink node. Specifically, we first do explicit (round-trip time based) maximum likelihood (ML) distance estimation, and obtain the corresponding Cramer Rao bound (CRB). We then invoke the (distance-based) impersonation detection framework proposed in (C1) for a UWA channel with colored noise and frequency-dependent pathloss.

Section II summarizes the prior art on security in UWASNs. But, to the best of authors' knowledge, {\it a systematic treatment of (network-wide) impersonation attack detection is missing in the existing literature on UWASNs}\footnote{The literature on physical layer security has mainly considered a very simplistic model consisting of only three nodes (Alice, Bob and Eve) so far \cite{Wang:ICST:2017},\cite{Xianbin:CommMag:2016}. This work, however, considers a more practical scenario where multiple Alice/sensor nodes report to a Bob/sink node. Therefore, we dub the proposed method as capable of doing network-wide impersonation detection.}.

\vspace{2mm}
{\bf Outline.}
The rest of this paper is organized as follows. Section II summarizes the selected related work. Section III presents the system model and the UWA channel model. Section IV proposes a novel, multiple-features based, two-step method for impersonation detection in an AWGN UWA channel. Section V obtains an explicit ML distance estimate and the corresponding CRB to carry out (distance-based) impersonation detection in a UWA channel with colored noise and frequency-dependent pathloss. Extensive simulation results are provided in section VI. Finally, Section VII concludes the paper.

\section{Related Work}

Security in UWASNs is a subject that has not yet received much attention by the researchers so far. There are a few review articles (\cite{Han:CommMag:2015}, \cite{Domingo:WiComm:2011}, \cite{Zhou:CCMC:2010}) and a vision paper \cite{Lal:JOE:2017} though which list various kinds of attacks which the malicious nodes could launch against the UWASNs, and provide their own take on design of futuristic secure UWASNs. The articles \cite{Han:CommMag:2015,Domingo:WiComm:2011,Zhou:CCMC:2010,Lal:JOE:2017} all admit that the security needs of UWSANs have not been addressed to full extent, i.e., there are many kinds of potential attacks (e.g. impersonation attack) for which no prevention/counter mechanisms have been reported in the literature. Nevertheless, the prior art on security in UWASNs is briefly summarized below.

The works in \cite{DiniCC:ISCC:2011},\cite{DiniND:ISCC:2011},\cite{Spaccini:OCEANS:2015} provide cryptographic solutions to address the security needs of UWASNs. The authors of \cite{DiniCC:ISCC:2011} consider both eavesdropping attack and the impersonation attack by the malicious node(s), and counter them by pre-distributing to the UWASN members a group key (which sensor nodes use to broadcast their sensed data to the group members) and a session key (which the sensor nodes use to send data to the sink node), while the sink node does the key management (e.g., the key generation, key updating, etc.). In \cite{DiniND:ISCC:2011}, the same authors extend a well-known network discovery protocol (where sensor nodes discover their neighbor to develop routing tables), the so-called FLOOD protocol, to protect the UWASN from the spoofing (impersonation) attacks and denial-of-service attacks by intruders during the network discovery phase. Specifically, the authors of \cite{DiniND:ISCC:2011} recommend that each UWASN node should be provided a link key table (a link key is the pairwise agreement/key between the two neighboring nodes). Moreover, the neighboring nodes form the clusters (a cluster is one collision domain) whereby all the cluster members share a cluster key to communicate with each other. Ateniese et al. \cite{Spaccini:OCEANS:2015} present various cryptographic solutions for message encryption and authentication, i.e., generation of (block cipher based) symmetric keys, and (elliptic curves based) asymmetric keys.

The works in \cite{Goetz:IWUWN:2011},\cite{Zuba:SCN:2015},\cite{Xiao:Globecom:2015} all consider jamming attacks on UWASNs by active (and aggressive) intruders. The authors of \cite{Goetz:IWUWN:2011} propose to route the sensed data to the sink node(s) via multiple paths (the so-called restricted flooding), which makes the system jamming-resilient. Zuba et al. \cite{Zuba:SCN:2015} conduct real-time jamming experiments with commercial (Benthos) acoustic OFDM modems in Mansfield Hollow Lake (in Mansfield, CT, USA) to demonstrate that jamming attacks could easily lead to denial of service predicament in UWASNs. Xiao et al. \cite{Xiao:Globecom:2015} utilize the tools from game theory to formulate the hostile interaction between jammers and UWASN nodes as a jamming game; the authors provide closed-form expressions for the Nash equilibrium when all the underwater channels are known. For the dynamic/uncertain underwater environments (when channels are not known),  Xiao et al. \cite{Xiao:Globecom:2015} utilizes a reinforcement learning-based power control scheme to prevent the jamming attacks.

The works in \cite{Dai:Sensors:2016}, \cite{Huang:SJ:2016} consider passive eavesdropping attacks by a malicious node Eve. In \cite{Dai:Sensors:2016}, authors consider a 2-D region (a disk) which consists of multiple UWASN nodes (and one Eve node) distributed according to a Poisson point process. The authors then utilize tools from stochastic geometry to compute the probability that the eavesdropper is able to intercept the communication ongoing within the network, and show that the probability of interception decreases as more and more legitimate nodes fall outside the critical region around the Eve. \cite{Huang:SJ:2016} considers a one-way, secure communication problem where a node Alice transmits to another node Bob (in the presence of an Eve node); Huang et al. \cite{Huang:SJ:2016} propose that the Bob node exploits the block transmissions nature and large propagation delays of the acoustic channel to send out a jamming signal which interferes with the Alice's signal received at Eve, thus maximizing the secrecy capacity of the acoustic channel. 

The works in \cite{Liu:ICSP:2008}, \cite{Huang:TWC:2016} study the problem of shared secret keys generation between a legitimate node pair by exploiting the physical-layer characteristics of the acoustic channel. To this end, Liu et al. \cite{Liu:ICSP:2008} exploit the amplitude (i.e., received signal strength) of (reciprocal) time-varying, multipath, acoustic channel as the source of common randomness, followed by a fuzzy information reconciliation system (to remove the inconsistencies between the keys generated by the two nodes). Huang et al. \cite{Huang:TWC:2016}, on the other hand, exploit the channel frequency response of the acoustic channel to generate the shared secret keys. \cite{Xu:TMC:2018} proposes SenseVault, a three-tier authentication framework to systematically generate (and update) cryptographic hash-based secret keys to authenticate the inter-cluster and intra-cluster UWASN nodes.
  
In short, to the best of authors' knowledge, {\it the problem of impersonation detection in UWASNs has not been reported in the literature yet}. On a side note, many experimental works have been reported in the literature on wireless sensor networks which attempt to do border surveillance and intrusion detection by deploying sensor nodes either over-the-ground or underwater, along the border (see the survey article \cite{Felemban:IJCNS:2013}). We note, however, that the works summarized in \cite{Felemban:IJCNS:2013} address the problem of an aggressive intruder (who is not interested to hide itself), while this work considers the scenario of a clever impersonator who aims to inject false data into the system while staying undetected. 

\section{System Model \& Channel Model}

\subsection{System Model}

\begin{figure}
\begin{center}
\includegraphics[width=3.4in, height=2.2in]{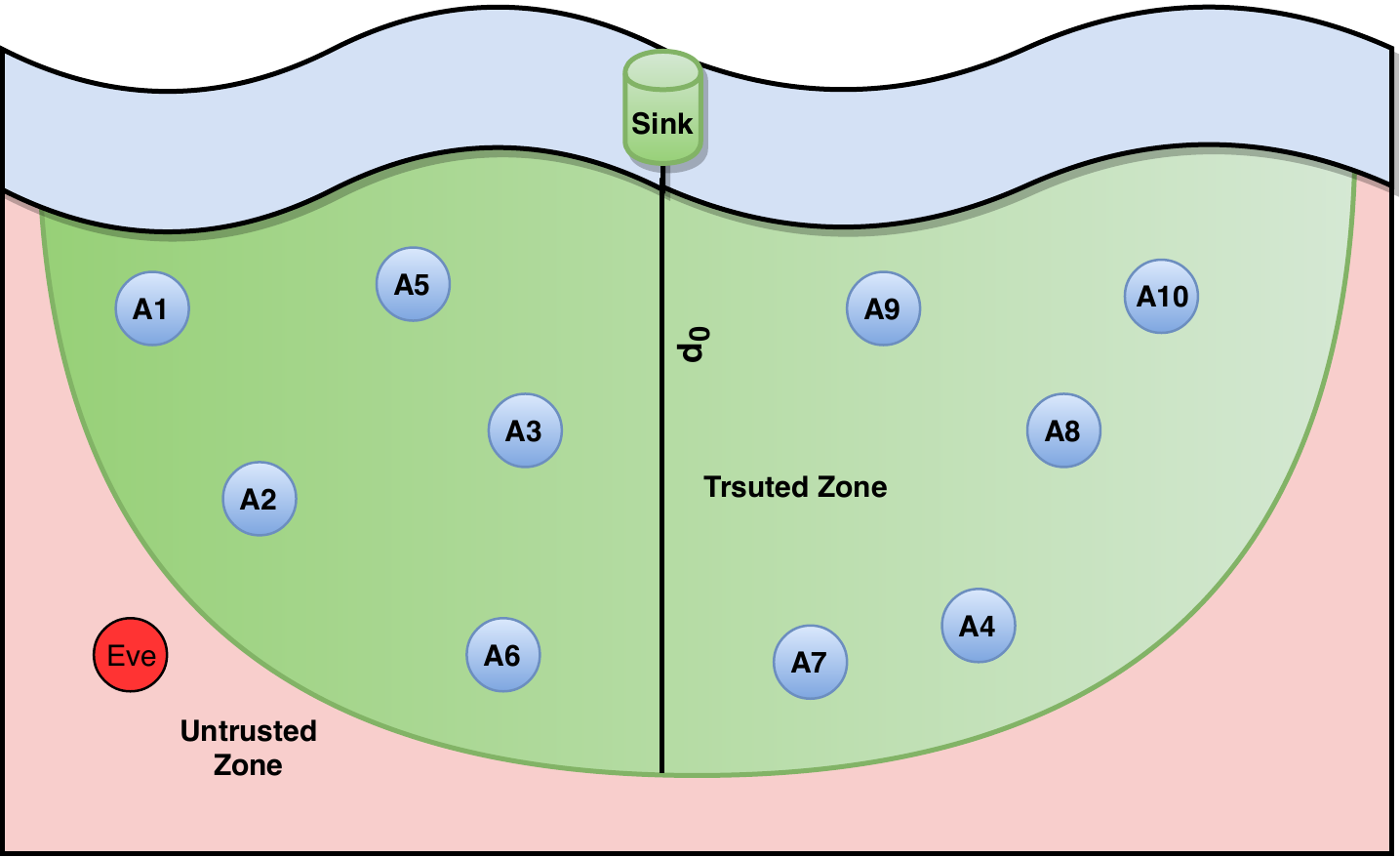}
\caption{An illustration of the system model using an example topology with $M=10$ sensor nodes, and an Eve node.}
\label{fig:system-model}
\end{center}
\end{figure}

We consider a UWASN comprising $M$ legitimate underwater sensor nodes (the so-called Alice nodes $\{A_i\}_{i=1}^{M}$) which report their sensed data to a sink node on the water surface (see Fig. \ref{fig:system-model}). The sensor nodes are deployed randomly (according to uniform distribution) on a vertical half-disc (the so-called trusted zone) according to a 2D geometry. All the nodes in the considered system model constitute one collision domain, i.e. the UWASN under consideration is a {\it single-hop system} whereby each sensor (Alice) node could send its sensed data {\it directly} to the sink node. The shared reporting channel is time-slotted; the sensor nodes access the reporting channel in a TDMA fashion (and thus, there are no collisions). The ongoing communication on the reporting channel is at risk of impersonation attack by a malicious node Eve present nearby. This work considers an attack scenario whereby the Eve is in active (but stealth) mode, i.e., Eve attempts to impersonate some sensor (Alice) node before the sink node so as to inject some malicious data into the system. 
We further assume the following: A1) All the nodes ($M$ legitimate nodes, the sink node as well as the impersonator Eve) are stationary; A2) Eve faithfully follows the communication protocol dictated by the sink node (to be described in the next section) in order to stay undetected; A3) The shared reporting UWA channel is memoryless\footnote{One example scenario of a memoryless channel is when the UWASN is deployed in deep waters, and the shared reporting channel has a small range-to-depth ratio (and thus a small range). Furthermore, the reporting channel is narrow-band (and thus low-rate), and vertical (and thus multipath reflections are negligible). This reporting channel then acts as a line-of-sight link which is noise-limited only \cite{Milica:JOE:1996}, \cite{Baggeroer:JOE:2000}.} (i.e., multipath is negligible); A4) The positions of the legitimate nodes $\{A_i\}_{i=1}^{M}$ are known to the sink node in advance\footnote{This is inline with the previous literature on impersonation attack detection at the physical layer\cite{Xianbin:CommMag:2016},\cite{Mahboob:Globecom:2014}.}.

\subsection{The UWA Channel with Colored Noise and Frequency-dependent Pathloss}
Two main attributes of the UWA channel degrading the performance of UWASNs are colored ambient noise, and frequency-dependent pathloss. Denote by $PL(d,f)$ the frequency-dependent pathloss between a transmit acoustic device and a receive acoustic device separated by distance $d$, and operating on frequency $f$. Then, $PL(d,f)$ is given (in dB scale) as \cite{Milica:MCCR:2007}:
\begin{align}
\label{eq:pl}
PL(d,f)_{dB} = \nu 10\log d + d \alpha(f)_{dB} 
\end{align}
where $\nu$ is the so-called spreading factor, while $\alpha(f)$ is the coefficient of absorption, given as \cite{Milica:MCCR:2007}: 
\begin{align}
\alpha(f)_{dB}=\frac{0.11            
f^2}{1+ f^2}+\frac{44f^2}{4100+f^2}+2.75\times 10^{-4}f^2+0.003 
\end{align}
Let $N(f)$ denote the power spectral density (PSD) of the frequency-dependent ambient noise (comprising of noise contributions from turbulence, shipping, waves, and thermal noise). Then, $N(f)$ is given (in dB scale) as \cite{Milica:MCCR:2007}:  
\begin{align}
\label{eq:noise}
N(f)_{dB} \approx N_1- \zeta 10\log f  
\end{align} 
where $N_1$ and  $\zeta$ are the experimental constants. Note that the above approximation of the PSD $N(f)$ of ambient noise holds for frequency range ($1-100$) kHz only \cite{Milica:MCCR:2007}.

\section{Impersonation Detection and Transmitter Identification in AWGN UWA Channel}
As briefly explained earlier, impersonation detection is a systematic framework to verify (at the physical layer) the identity of the sender node so as to detect-then-reject the data coming from the (stealth) impersonator node in order to maintain data integrity of the system. For this section, we make the following additional assumptions: B1) The shared reporting UWA channel is AWGN\footnote{That is, the colored noise inherent to the system has been transformed into white noise by means of a pre-whitening filter at the sink node \cite{Berger:JASA:2010}.}; B2) The noisy estimates of the distance and AoA (and thus, position) of the channel occupant are available at the sink node\footnote{For example, the distance could be estimated using two-way ranging based localization schemes \cite{Mouftah:CST:2011},\cite{Poor:TWC:2010},\cite{Tan:JOE:2011}, while the work \cite{Li:TSP:2011} (and other works by the same authors) describes various ways to estimate the AoA.}. Note that both assumptions B1), B2) are relaxed in the next section where we obtain an explicit ML distance estimate and the corresponding CRB to carry out (distance-based) impersonation detection in a UWA channel with colored noise and frequency-dependent pathloss.

The proposed method consists of two steps, which work together to carry out impersonation detection and transmitter identification. The first (second) step works under the assumption that the Eve node is outside (inside) the so-called trusted zone. The first step consists of a distance bounding test, while the second step consists of three outlier detection tests.

\subsection{Step 1: Distance-bounding test}
This step is inspired by the proximity-based authentication techniques (which trust those transmit nodes only that are in the close proximity) in the radio-frequency identification systems \cite{Kuhn:ICSPEACN:2005}, and the works on border intrusion detection \cite{Felemban:IJCNS:2013}. This step assumes that Eve, being a clever impersonator, wants to remain undetected; therefore, it remains outside the trusted zone. As otherwise, if Eve enters the trusted zone, it might be detected by the system due to the on-board proximity sensors of the Alice node(s) \cite{Felemban:IJCNS:2013}. 

{\bf The trusted zone.} As a first layer of defense against the potential intrusion, the system relies upon the so-called trusted zone, a pre-defined geographic region around the sink node (i.e., a virtual fence). Specifically, this work considers a trusted zone which is a half-disc\footnote{The trusted zone is a half-disc because under the distance bounding protocol, the sink node trusts the transmissions from the sender nodes which are less than $d_0$ distance away and vice versa.} of radius $d_0$ when the sink node is placed at the origin (see Fig. \ref{fig:system-model}). Under step 1, all the nodes inside the trusted zone (the half-disc) are considered to be legitimate nodes, while all the nodes outside the trusted zone are considered to be malicious/other nodes. 

\begin{figure}
\includegraphics[width=3in, height=1.3in]{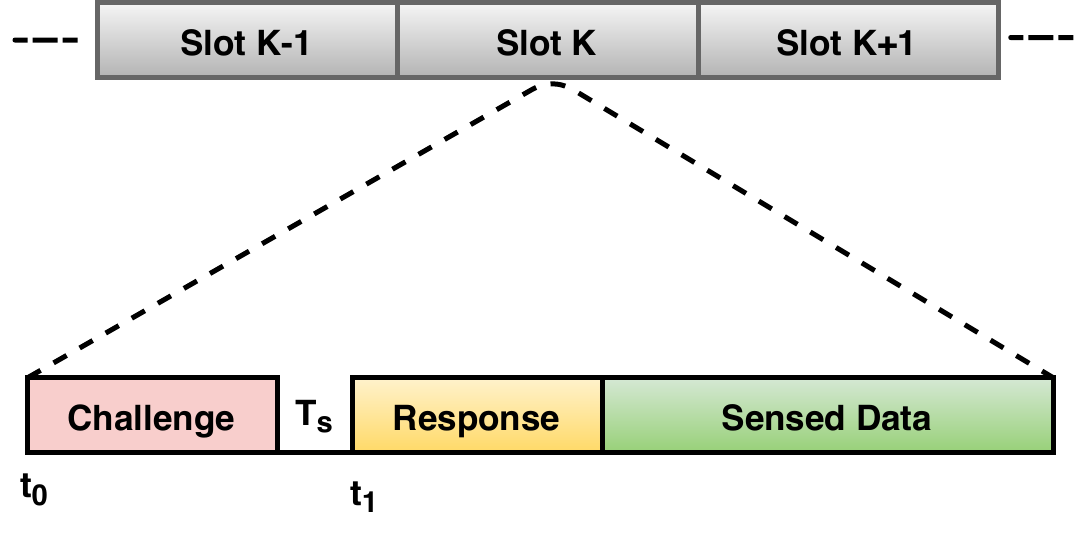}
\caption{Timeline of the TDMA reporting UWA channel}
\label{fig:timeline}
\end{figure}

{\bf The distance-bounding protocol.} Whenever the sink node receives some data on the shared reporting channel, it has to authenticate the sender of the data. As for the step 1, the sink node needs to estimate whether the sender node is inside the trusted zone or outside it. To this end, this work exploits the distance bounding protocol which works as follows. In the beginning of every time-slot, the sink node broadcasts a ``challenge message" (see Fig. \ref{fig:timeline}) which serves two purposes: i) it announces the beginning of the current time-slot to all the UWASN nodes, ii) it asks the channel claimant of the upcoming time-slot to prove its identity via transmission of a ``response message". This two-way communication constitutes the challenge-response based distance-bounding protocol \cite{Kuhn:ICSPEACN:2005}. Specifically, each challenge message from the sink node contains a (different) pseudo-noise (PN) sequence. The channel claimant node is required to echo back the PN sequence (after a delay of $T_s$) by putting it in its response message\footnote{$T_s$ arises due to hardware limitations of a wireless/acoustic device to switch from receive mode to transmit mode. In this work, the sink node pre-broadcasts a value for $T_s$ (larger than the typical switching delays), which the channel claimant must abide by.}.

{\bf Distance as transmit device fingerprint.}
Under distance-bounding protocol, the sink node needs to estimate the distance of the channel claimant from itself during every time-slot. To this end, the sink node obtains the distance estimate via (the challenge-response based) two-way ranging method. That is, the sink node marks the time instant $t_0$ of beginning of the challenge message; and a while later, estimates the time of arrival (ToA) $t_1$ of the received response message by correlating the received noisy PN sequence against the stored copy of the same PN sequence, and marking the time instant where the correlation is maximum. The sink node then translates the estimate of the round-trip time (RTT) $t_1-t_0$ to a distance estimate as $v(t_1-t_0)$ where $v$ is the speed of sound wave underwater.

\vspace{2mm}
{\bf \textit{Test 1}: The distance bounding test.} During the $k$-th time-slot, after computing the unbiased distance estimate $z(k)=\hat{d}$, the sink node implements the test 1 as the following binary hypothesis test:

\begin{equation}
	\label{eq:H0H1-db}
	 \begin{cases} H_0 (\text{sender is in trusted zone}): & {z}(k)=d_i+n_d(k) \\ 
                                 H_1 (\text{sender is in untrusted zone}): & {z}(k)=d_E+n_d(k) \end{cases}
\end{equation}
where $d_i$ ($d_E$) is the distance of the $A_i$ (Eve) node from the sink node, and $n_d\sim \mathcal{N}(0,\sigma_d^2)$ is the estimation error. Since all the Alice nodes are deployed within the trusted zone, the binary hypothesis (BH) test in Eq. (\ref{eq:H0H1-db}) translates to the following test:
\begin{align} 
\label{eq:test-db}
z(k) \gtrless_{H_0}^{H_1} {d_0}
\end{align}
The test 1 depicted in Eq. (\ref{eq:test-db}) approves the transmission from a sender node if the sender node is less than $d_0$ distance away from the sink node and vice versa.

\vspace{2mm}
{\bf Performance of the test 1.} The BH test of Eq. (\ref{eq:test-db}) will incur two kinds of errors: false alarm (i.e., misclassifying some $A_i$ as Eve), and missed detection (i.e., misclassifying Eve as some $A_i$). The probabilities for the both error events are as follows. The probability of false alarm is given as:

\begin{equation}
P_{fa}=\sum_{i=1}^M Pr(z(k)>d_0|A_i)\pi(i)
\end{equation} 
where $z(k)|A_i \sim \mathcal{N}(d_i,\sigma_d^2)$; $\pi(i)$ is the prior probability that the $i$-th Alice node $A_i$ becomes the channel occupant during the $k$-th time-slot. This work considers the case of equal priors, i.e., $\pi(i)=\frac{1}{(M+1)}$. Then, 
\begin{equation}
\label{eq:pfa1}
P_{fa}=\frac{1}{(M+1)} \sum_{i=1}^M Q(\frac{d_0-d_i}{\sigma_d})
\end{equation} 
where $Q(x)=\int_{x}^{\infty } \frac{e^{\frac{-t^2}{2}}}{\sqrt{2\pi}} \ dt$ is the standard $Q$-function.

Next, the probability of missed detection (the success rate of Eve) is given as:
\begin{equation}
P_{md}=Pr(z(k)<d_0|E)\pi(E)
\end{equation}
where $z(k)|E \sim \mathcal{N}(d_E,\sigma_d^2)$; $\pi(E)$ is the prior probability that Eve node becomes the channel occupant during the $k$-th time-slot. Since $P_{md}$ is a random variable (RV) (because the unknown distance $d_E$ is an RV), we compute its expected value $\bar{P}_{md}:={E}(P_{md})$ as follows:
\begin{equation}
\label{eq:pmd1}
\bar{P}_{md}=\frac{1}{(M+1)}\bigg( 1 - \int_{d_0+\epsilon}^{kd_0} \eta \; Q(\frac{d_0-d_E}{\sigma_d}) dd_E \bigg)
\end{equation}
where we have assumed that $d_E\sim U(d_0+\epsilon,kd_0)$; $\epsilon >0$ is a small number and $k> 1$, and $\eta=\frac{1}{d_0(k-1)-\epsilon}$ is the probability density function (PDF) of $d_E$. 

\vspace{2mm}
{\bf Remark 1.} Despite its simplicity, the main strength of the distance-bounding protocol is that Eve cannot deceive the sink node by making her believe that Eve is a trusted node which lies inside the trusted zone. This is because Eve cannot tamper with the speed of acoustic waves underwater to make $d_E$ appear lesser than $d_0$ before the sink node. On the other hand, Eve could indeed make $d_E$ appear greater than $d_0$ by delaying the response message (beyond the value $T_s$ suggested by the protocol, see Fig. \ref{fig:timeline}). It is noted, however, that such tampering will not favor Eve, as the sole intent of distance bounding protocol is to reject network access requests (and/or data) from the transmit nodes that are $>d_0$ distance away. On a different note, if Eve tries to send a response message (containing the malicious payload) before the challenge message is sent by the sink node, Eve will be detected due to two reasons: i) Eve's transmission could collide with the transmission of some (scheduled) Alice node from the previous slot; ii) Eve does not know the PN sequence the sink node has sent in its latest challenge message. 

\subsection{Step 2: Outlier detection tests}
This step addresses the scenario when Eve is potentially present within the trusted zone (e.g., because the on-board proximity sensors of the nearby Alice node(s) within the trusted zone were defunct). In such situation, step 1 fails to detect any impersonation attack. Therefore, (as the second layer of defense) the sink node implements the step 2, which utilizes the AoA and position as additional device fingerprints. 

\begin{figure}
\includegraphics[width=9cm, height=5cm]{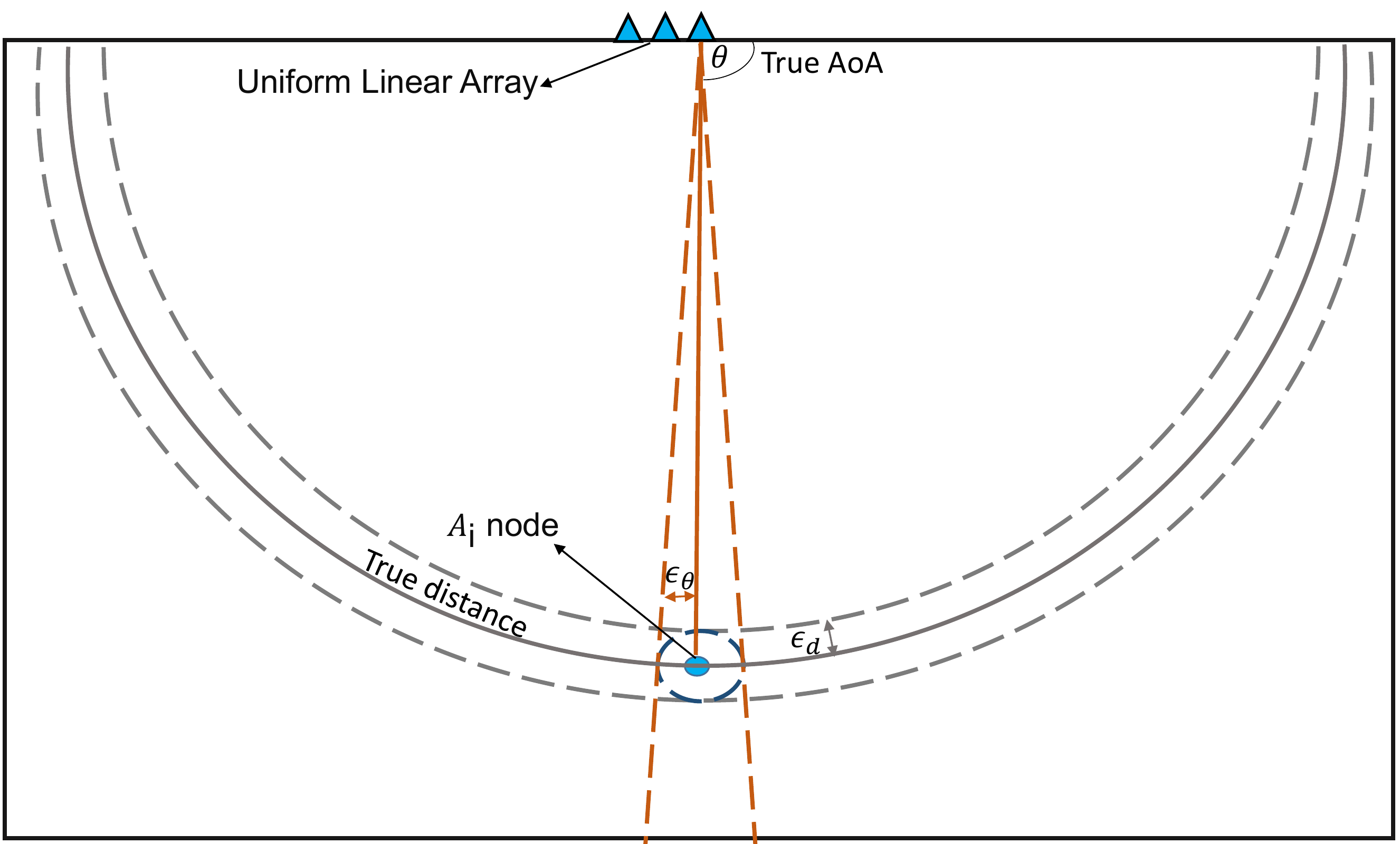}
\caption{The proximity regions of the three tests in step 2 (The sink node is shown to be equipped with a ULA containing three hydrophones.)}
\label{fig:PRs}
\end{figure}

\vspace{1mm}
{\bf AoA and Position as transmit device fingerprints.} When Eve is inside the trusted zone, the distance alone ceases to be effective as the fingerprint of the transmit node(s). This is because in this case $P(|d_i-d_E|<\xi)>0$ for some $i$, $i=1,...,M$ ($\xi$ is a small number). Therefore, to resolve the situation when $d_E$ is very similar to $d_i$ (for some $i$), this step incorporates the AoA as an additional fingerprint of the transmit device. Let 
\begin{equation}
y(k)=\hat{\theta}(k)=\theta+n_{\theta}(k)
\end{equation}

where $y(k)$ represents the AoA measurement during the $k$-th time-slot; $\theta$ is the true AoA of the transmit node\footnote{Assuming that the uniform linear array (ULA) of hydrophones at the sink node is horizontally placed on the water surface (along the positive x-axis), the AoA is the angle made by a sensor node from positive x-axis in counter clockwise direction (see Fig. \ref{fig:PRs}).}; $n_{\theta}(k) \sim \mathcal{N}(0,\sigma_{\theta}^2)$ is the estimation error\footnote{\cite{Li:TSP:2011} describes various methods to estimate the AoA in UWASNs.}. Then, $\hat{p}(k)=z(k)\exp{(jy(k))}$ is the (derived) position estimate of the transmit node, obtained by the sink node during the $k$-th time-slot. In other words, the sink node performs a ranging-based source localization \cite{Tan:JOE:2011} and then the location estimate is used as fingerprint of the transmit device.

This work assumes that the positions of the legitimate nodes (a.k.a the ground truth) are known to the sink node in advance. In other words, $\mathbf{d}=\{d_1,...,d_M\}^T$, $\mathbf{\Theta}=\{\theta_1,...,\theta_M\}^T$; and therefore, $\mathbf{p}=\{p_1,...,p_M\}^T$ (where $p_i=d_i\exp{(j\theta_i)}$) are available at the sink node. Then, for each of three fingerprints, the step 2 consists of an interplay between two kinds of sub-tests: a maximum likelihood (ML) hypothesis test followed by another BH test. As a by-product, the step 2 enables the sink node to perform transmitter identification (for the no impersonation case) as well.

{\bf \textit{Test 2(a)}: Position based test.} The ML sub-test works as follows: 

\begin{equation} 
i_p^*=\underset{1\leq i \leq M}{\arg\max} \quad {f}_{\hat{P}|A_i}(\hat{p}(k)) 
\end{equation} 
where $f_{\hat{P}|A_i}$ is the PDF of $\hat{P}|A_i$. Essentially, the ML test returns the index $i_p^*$ that maximizes the likelihood value $f_{\hat{P}|A_i^*}$, given the noisy observation $\hat{p}(k)$. However, we note that the closed-form expression for the pdf $f_{\hat{P}|A_i}$ $\forall i$ is hard to derive. Therefore, we propose an alternative (sub-optimal) approach, the nearest-neighbour test. Let: 

\begin{equation} 
\label{eq:ML-P}
(J^*,i_p^*) = \underset{i}{\min} \quad ||\hat{p}(k)-p_i||_2 
\end{equation}

Note that due to lack of prior knowledge about $p_E$ (the position of Eve), the ML test only solves the transmitter identification problem (for Alice nodes, for the no impersonation case). For impersonation detection, one needs to define another binary hypothesis test which works as follows: if $\underset{i}{\min} ||\hat{p}(k)-p_i||_2 > \epsilon_p$, then outlier/Eve is detected; else, $A_i$ from the ML test is declared to be the sender of the data ($\epsilon_p$ is a small threshold, a design parameter). Equivalently, the BH sub-test is:

\begin{equation}
	\label{eq:H0H1-P}
	 \begin{cases} H_0 (\text{no impersonation}): & J^*=\underset{i}{\min} ||\hat{p}(k)-p_i||_2 < \epsilon_p \\ 
                                 H_1 (\text{impersonation}): & J^*=\underset{i}{\min} ||\hat{p}(k)-p_i||_2 > \epsilon_p \end{cases}
\end{equation}
The BH test in Eq. (\ref{eq:H0H1-P}) can be re-written as:
\begin{align} 
\label{eq:test-P}
J^* \gtrless_{H_0}^{H_1} {\epsilon_p}
\end{align}
The test in Eq. (\ref{eq:test-P}) approves the transmission from a sender node only if the position estimate $\hat{p}(k)$ of the sender node lies within the ball (around some point $p_i$, $i=1,...,M$) of radius $\epsilon_p$ and vice versa.

\vspace{2mm}
{\bf \textit{Test 2(b)}: Distance based test.} 
The ML (equivalently, the nearest-neighbour) sub-test works as follows:  
\begin{equation} 
\label{eq:ML-d}
(K^*,i_d^*) = \underset{i}{\min} \quad |z-d_i|
\end{equation}
Next, the BH sub-test works as follows: 
\begin{align} 
\label{eq:test-d}
K^* \gtrless_{H_0}^{H_1} {\epsilon_d}
\end{align}
where $\epsilon_d$ is a small threshold, a design parameter.

\vspace{2mm}
{\bf \textit{Test 2(c)}: AoA based test.}
The ML sub-test works as follows: 
\begin{equation} 
\label{eq:ML-aoa}
(L^*,i_\theta^*) = \underset{i}{\min} \quad |y-\theta_i|
\end{equation}

Next, the BH sub-test works as follows: 
\begin{align} 
\label{eq:test-aoa}
L^* \gtrless_{H_0}^{H_1} {\epsilon_{\theta}}
\end{align}
where $\epsilon_\theta$ is a small threshold, a design parameter.

\vspace{2mm}
{\bf Remark 2.} 
The closed-form expressions for the two error probabilities (i.e., $P_{fa}$ and $P_{md}$) could not be derived for the test 2(a) since the PDF of the test statistic $J^*$ in Eq. (\ref{eq:test-P}) is not straightforward to obtain. However, Section VI shares extensive simulation results which shed light on the performance of the tests 2(a), 2(b), 2(c) as well as the fusion rules (discussed below).

\vspace{2mm}
{\bf Performance of test 2(b).} 
The two error probabilities for test 2(b) are:
\begin{equation}
\label{eq:pfa2}
\begin{split}
P_{fa}^{(d)}&=P(K^*>\epsilon_d|H_0) \\
&=\frac{1}{(M+1)} \sum_{i=1}^M 2Q(\frac{\epsilon_d}{\sigma_d})=\frac{2M}{(M+1)}Q(\frac{\epsilon_d}{\sigma_d})
\end{split}
\end{equation}
and $\bar{P}_{md}^{(d)}:={E}(P_{md}^{(d)})$ is as follows:
\begin{equation}
\label{eq:pmd2}
\begin{split}
\bar{P}_{md}^{(d)} & = E(P(K^*<\epsilon_d|H_1)) \\
& =\frac{1}{(M+1)(kd_0-d_{min})}. \\
& \bigg( \int_{d_{min}}^{kd_0} \sum_{i=1}^M Q(\frac{d_i-\epsilon_d-d_E}{\sigma_d})-Q(\frac{d_i+\epsilon_d-d_E}{\sigma_d}) dd_E \bigg)
\end{split}
\end{equation}
where we have assumed that the unknown distance $d_E\sim U(d_{min},kd_0)$. 

The expressions for $P_{fa}^{(\theta)}$ and $P_{md}^{(\theta)}$ for test 2(c) could be obtained in a similar way; and therefore, are omitted for the sake of brevity. 

\vspace{2mm}
{\bf Remark 3.}
Each of the tests 2(a), 2(b) \& 2(c) checks whether or not the noisy measurement of sender's fingerprint is within the so-called proximity region (PR) of any of the legitimate (Alice) nodes and decides accordingly. The PR, by definition, is a small region around the true value of each fingerprint, which represents the estimation errors. The PR is a half-ring (of width $2\epsilon_d$ meters) for the distance test, a cone (of width $2\epsilon_\theta$ degrees) for the AoA test, and a circle (of radius $\epsilon_p$ square meters) for the position test (see Fig. \ref{fig:PRs}). As Section VI will demonstrate, various levels of performance could be obtained by varying the size of the PR (or, equivalently, by varying the comparison thresholds $\epsilon_p$, $\epsilon_d$ \& $\epsilon_\theta$).

\vspace{1mm}
\subsection{Impersonation Detection}
To detect the potential impersonation, first the individual binary decisions---\textit{impersonation} or \textit{no impersonation}---of all the three tests in the second step are fused together. Then, the fusion outcome is further fused with the binary decision from the first step to generate the ultimate binary decision.

{\bf The decision fusion of tests 2(a), 2(b) and 2(c).} 
The individual decisions of tests 2(a), 2(b), 2(c) are fused via i) AND rule, ii) OR rule, iii) majority voting (MV) rule. Specifically, the AND (OR) rule is pessimistic (optimistic), i.e., a sender node is authenticated only if all (any one out of) the three tests decide $H_0$. The AND (OR) rule strives to minimize $P_{md}$ ($P_{fa}$).

\vspace{1mm}
{\bf The decision fusion of step 1 and step 2.}
When Eve is inside the trusted zone, step 1 is not helpful; therefore, only the outcome of step 2 should count to decide about the potential impersonation. On the other hand, when Eve is outside the trusted zone, the outcome of step 1 is equally helpful. To take into account both situations, this work applies the (pessimistic) AND rule to fuse the individual decisions made by step 1 \& step 2 (which minimizes the ultimate probability of missed detection even further).

\vspace{1mm}
\subsection{Transmitter Identification} 
When both steps (step 1 and step 2) declare $H_0$, i.e., no impersonation, then $i^*=MV(i_p^*,i_d^*,i_{\theta}^*)$ works as the transmit identifier. In this situation, the probability of misclassification error is given as: 
\begin{equation}
P_e = \sum_{i=1}^M P_{e|i}\pi(i)
\end{equation}
where $P_{e|i}=P(\text{sink decides }A_j|A_i\text{ was the sender})$. For the distance based test (test 2(b)), $P_{e|i}$ is given as:
\begin{equation}
\label{eq:pmc}
P_{e|i}^{(d)}=1-\bigg( Q(\frac{\tilde{d}_{l,i}-\tilde{d}_i}{\sigma_d}) - Q(\frac{\tilde{d}_{u,i}-\tilde{d}_i}{\sigma_d}) \bigg)
\end{equation}
where $\tilde{d}_{l,i}=\frac{\tilde{d}_{i-1}+\tilde{d}_i}{2}$, $\tilde{d}_{u,i}=\frac{\tilde{d}_{i}+\tilde{d}_{i+1}}{2}$. Additionally, $\mathbf{\tilde{d}}=\{\tilde{d}_{1},...,\tilde{d}_{M}\}=\text{sort}(\mathbf{d})$ where sort(.) operation sorts a vector in an increasing order. For the boundary cases, e.g., $i=1, i=M$, $\tilde{d}_{l,1}=d_{min}$, $\tilde{d}_{l,M}=d_{0}$ respectively.

A similar expression exists for the misclassification error $P_{e|i}^{(\theta)}$ for the AoA-based test (test 2(c)) which is omitted for the sake of brevity. 

\begin{figure}
\includegraphics[width=9cm, height=7cm]{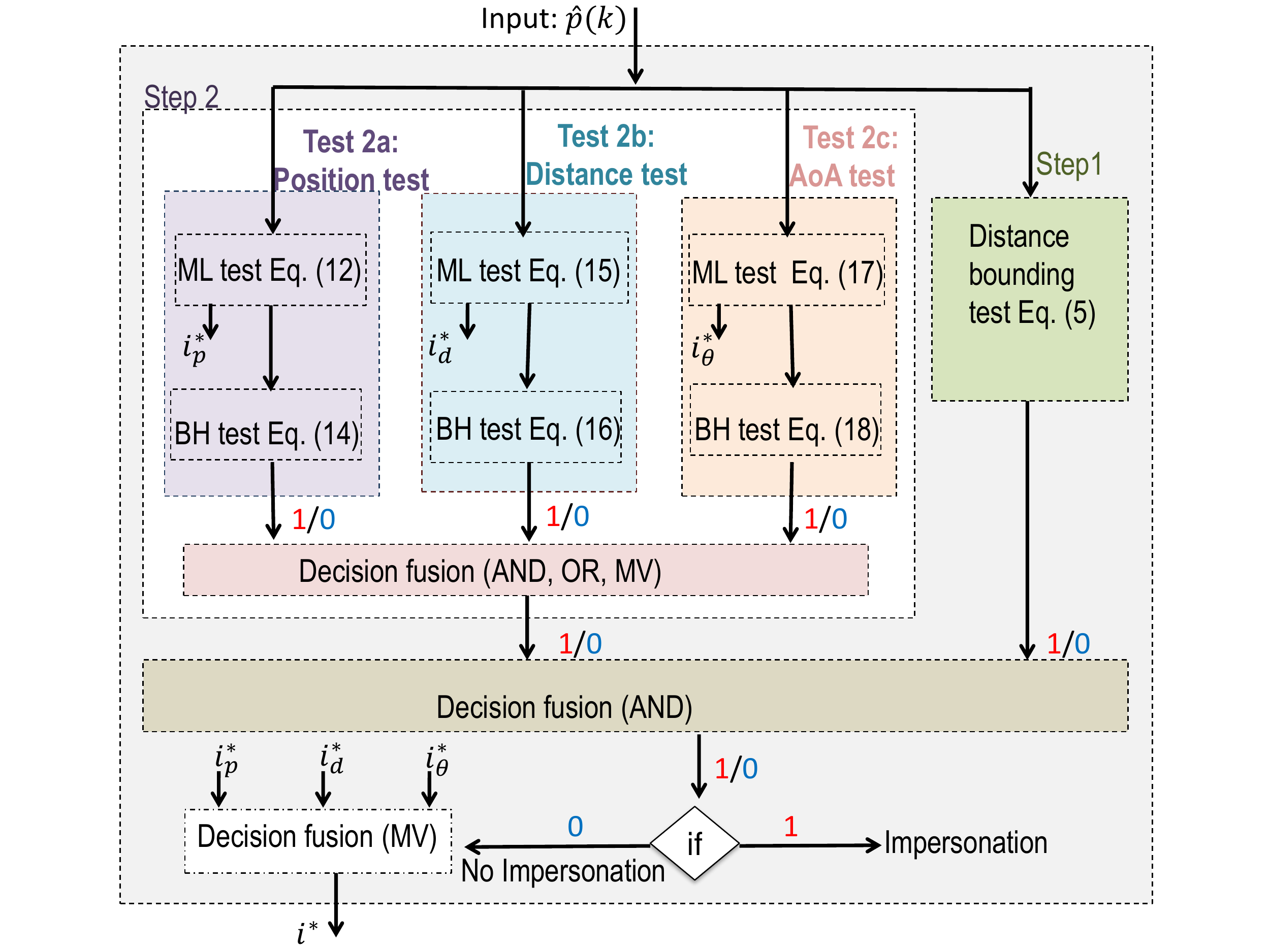}
\caption{The flow chart of the proposed method for Impersonation detection \& Transmitter identification}
\label{fig:flowchart}
\end{figure}

The algorithmic implementation of the proposed method has been summarized in Algorithm 1, 
while Fig. \ref{fig:flowchart} provides a graphical summary.

\begin{algorithm}
\label{algo:IDandTXID}
    \SetKwInOut{Input}{Input}
    \SetKwInOut{Output}{Output}

    \Input{$\hat{p}(k)=z(k)\exp{(jy(k))}$}
    \Output{$b$, $i^*$ \textcolor{blue}{// $i^*$ is the index of the sender node; $b=1$ ($b=0$) implies (no) impersonation. }}
    \Parameter{ $\mathbf{p}$, $\mathbf{d}$, $\mathbf{\Theta}$, $d_0$, $\epsilon_p$, $\epsilon_d$, $\epsilon_\theta$, $k$ }
     \underline{Step 1: Distance bounding test}:\\
     implement the BH test in Eq. (\ref{eq:test-db}) and return binary decision\\
     \underline{Step 2: Outlier detection tests}:\\
      implement the ML tests in Eq. (\ref{eq:ML-P}), Eq. (\ref{eq:ML-d}), Eq. (\ref{eq:ML-aoa}) to return $J^*,K^*,L^*$ and $i_p^*,i_d^*,i_{\theta}^*$ \\
      implement the BH tests in Eq. (\ref{eq:test-P}), Eq. (\ref{eq:test-d}), Eq. (\ref{eq:test-aoa}) to return binary decisions for each test\\ 
      \underline{Fusion of tests in step 2}:\\
      apply AND, OR, MV rules to fuse the individual decisions by tests in Eq. (\ref{eq:test-P}), Eq. (\ref{eq:test-d}), Eq. (\ref{eq:test-aoa})\\
      \underline{Impersonation detection:} \\
      apply AND rule to fuse the binary decisions by step 1 and step 2 \\
      \underline{Transmitter identification:}\\
      apply MV rule on $i_p^*,i_d^*,i_{\theta}^*$ to return the index $i^*$ when $H_0$ is decided
      
    \caption{The proposed method for Impersonation detection \& Transmitter identification}
\end{algorithm}

\section{Impersonation Detection and Transmitter Identification in UWA Channel with Colored Noise and Frequency-dependent Pathloss}

In this section, we first derive an explicit ML distance estimate and the corresponding CRB in a UWA channel with colored noise and frequency-dependent pathloss to relax the assumptions B1) and B2) made earlier in Section IV. We then carry out a single-feature (distance) based impersonation detection and transmitter identification by invoking Test 1 and Test 2(b) from Algorithm 1 proposed in Section IV. 

\subsection{Round-Trip Time/ToA based Distance Estimation}
\label{app:d_est} 

For a (colored) noise-limited, line-of-sight UWA channel that is exposed to frequency-dependent pathloss, the signal $y(t)$ received at the sink node is given as: $y(t)=\sqrt{P_R}s(t-t_1)+w(t)$ where $s(t)$ is the response message sent by channel claimant, and $t_1$ is the ToA to be estimated. $w(t)$ is the stationary Gaussian colored noise process with auto-correlation function $\mathcal{R}_w(\tau)=\mathcal{F}^{-1}\{N(f)\}$ with lag parameter $\tau$; $\mathcal{F}^{-1}\{.\}$ is the inverse Fourier transform operator; $N(f)$ is the PSD of the colored noise defined in Eq. (\ref{eq:noise}). Finally, $\sqrt{P_R}=\sqrt{\frac{P_T}{PL}}$ where $PL$ is the pathloss given in Eq. (\ref{eq:pl}), and $P_T$ is the fixed transmit power used by the channel claimant.

The equivalent discrete-time model for the signal received at the sink node is: $y(nT_{\mathcal{S}})=y[n]=\sqrt{P_R}s[n-t_1]+w[n]$ where $y[n]$ is the output of the receive filter, and ${T_\mathcal{S}} \le T_b/2$ (to avoid aliasing) is the sampling interval; $T_b$ is the bit duration. We assume that the sink node collects $Q$ samples during one slot. Then we can write: $\mathbf{y}=\mathbf{s}+\mathbf{w}$ where $\mathbf{y}=\{y[n]\}_{n=1}^Q$, $\mathbf{s}=\{\sqrt{P_R}s[n-t_1]\}_{n=1}^Q$, and $\mathbf{w}=\{w[n]\}_{n=1}^Q$. Then, under maximum likelihood (ML) estimation framework, the ToA estimate is the one which maximizes the (log of) joint (conditional) density:
\begin{equation}
\begin{split}
\hat{t}_{1}&= \arg \max_{t_1} \log f_\mathbf{y}(\mathbf{y}|t_1)  = \arg \max_{t_1} L(\mathbf{y};t_1) \\
&=  \arg \max_{t_1} \log \frac{1}{(2\pi)^{Q/2} |\mathbf{C}|^{1/2}}\exp{(-\frac{1}{2}(\mathbf{y}-\mathbf{s})^T \mathbf{C}^{-1} (\mathbf{y}-\mathbf{s}) )} \\
& = \arg \max_{t_1} \bigg[ -\log {(2\pi)^{Q/2} |\mathbf{C}|^{1/2}} - \frac{1}{2}(\mathbf{y}-\mathbf{s})^T \mathbf{C}^{-1} (\mathbf{y}-\mathbf{s}) \bigg]
\end{split}
\end{equation}
where $\mathbf{C}=E\{\mathbf{w}\mathbf{w}^T\}$ is $Q\times Q$ covariance matrix of $\mathbf{w}$, and $|.|$ represents the determinant of a matrix. Note that $\mathbf{w}\sim \mathcal{N}(\mathbf{0},\mathbf{C})$ where $\mathbf{0}$ is the vector (of appropriate size) of all zeros; $[\mathbf{C}]_{i,j}=\sigma^2[\check{\mathbf{C}}]_{i,j}$ where $[\check{\mathbf{C}}]_{i,j}=\mathcal{R}_w(\tau=|i-j|T_{\mathcal{S}})=\mathcal{R}_w[|i-j|]$, $0 \le [\check{\mathbf{C}}]_{i,j}\le 1$; $[\check{\mathbf{C}}]_{i,i}=\mathcal{R}_w(0)=1$ $\forall$ $i,j=1,..,Q$.

Discarding the irrelevant terms and rearranging, we have:
\begin{equation}
\label{eq:mlls}
\hat{t}_{1} = \arg \max_{t_1} L(\mathbf{y};t_1) = \arg \min_{t_1} \; (\mathbf{y}-\mathbf{s}(t_1))^T \mathbf{C}^{-1} (\mathbf{y}-\mathbf{s}(t_1)) 
\end{equation}
where the notation $\mathbf{s}(t_1)$ is used to highlight the dependence of $\mathbf{s}$ on $t_1$. Eq. (\ref{eq:mlls}) is indeed a matched filtering operation where that $t_1$ is chosen which maximizes (minimizes) the weighted inner product $\langle \mathbf{y},\mathbf{s} \rangle_{\mathbf{C}^{-1}}$ ($\langle \mathbf{y}-\mathbf{s} \rangle_{\mathbf{C}^{-1}}$) where $\langle \mathbf{a},\mathbf{b} \rangle_{\mathbf{D}}=\mathbf{a}^T \mathbf{D} \mathbf{b}$. In other words, the ML delay/ToA estimator block is simply a matched filter, or, the PN sequence correlator (which compares the received noisy signal against the delayed copies of the pre-stored clean PN sequence $s[n]$).

One can verify that $\frac{\partial L(\mathbf{y};t_1)}{\partial t_1} = 2\mathbf{\dot{s}}^T \mathbf{C}^{-1}(\mathbf{y}-\mathbf{s})$ where $\mathbf{\dot{s}}=\sqrt{P_R}\frac{\partial}{\partial t_1} \{ s[n-t_1]\}_{n=1}^Q$. Setting $\frac{\partial L(\mathbf{y};t_1)}{\partial t_1}=0$, we get a transcendental equation. Therefore, no closed-form expression exists for ML estimate of $t_1$, and we resort to Eq. (\ref{eq:mlls}) to compute $\hat{t}_1$ via exhaustive search by plugging $Q$ values of $t_1$, i.e., $t_1 \in \{n_0,...,n_0+Q-1\}$ (we take $n_0=1$ here). For this, $P_R$ is estimated as: $\hat{P}_R=\frac{1}{Q}\sum_{n=1}^Q (y[n])^2$.

Also, one can verify that: $\frac{\partial^2 L(\mathbf{y};t_1)}{\partial^2 t_1}=-4(\mathbf{\dot{s}}^T \mathbf{C}^{-1} \mathbf{\dot{s}} + \mathbf{\ddot{s}}^T \mathbf{C}^{-1} (\mathbf{{y}}-\mathbf{{s}}) )$, where $\mathbf{\ddot{s}}=\sqrt{P_R}\frac{\partial^2}{\partial^2 t_1} \{ s[n-t_1]\}_{n=1}^Q$. With this, the Cramer-Rao bound (CRB) for the ToA estimate is obtained as:
\begin{equation}
\begin{split}
\text{Var}(\hat{t}_1) &\ge \text{CRB}(\hat{t}_1) = -\frac{1}{E[\frac{\partial^2 L(\mathbf{y};t_1)}{\partial^2 t_1}]} = \frac{1}{E[(\frac{\partial L(\mathbf{y};t_1)}{\partial t_1})^2]} \\
=& \frac{\sigma^2}{ 4 P_R \mathbf{\dot{s}}^T \check{\mathbf{C}}^{-1} \mathbf{\dot{s}} } = \frac{\sigma^2 PL}{ 4 P_T \mathbf{\dot{s}}^T \check{\mathbf{C}}^{-1} \mathbf{\dot{s}} } 
\end{split}
\end{equation}
Since the proposed ML estimate satisfies regulatory conditions on $L(\mathbf{y};t_1)$ (i.e., the first two derivatives of $L(\mathbf{y};t_1)$ w.r.t. $t_1$ exist, and Fisher information $I(t_1)=-E[\frac{\partial^2 L(\mathbf{y};t_1)}{\partial^2 t_1}]$ is non-zero); therefore, (for large $Q$) it is asymptotically optimal, unbiased and Gaussian. In other words, $\hat{t}_1 \stackrel{a}{\sim} \mathcal{N}(t_1,I(t_1)^{-1})$. Thus, $\hat{t}_1$ is efficient, i.e., it meets the CRB. 

Having estimated the ToA $t_1$, the sink node then computes an estimate of the round-trip time (RTT) as follows: $\widehat{\Delta t}= \hat{t}_1-t_0$. Equivalently, $\widehat{\Delta t}=\widehat{\Delta t_p}+T_s$, where $T_s$ is the switching delay, and $\widehat{\Delta t_p}$ is the RTT with zero switching delay. With this, the sink node obtains the following distance estimate: 
\begin{equation}
\label{eq:dhat}
z=\hat{d}=v\frac{\widehat{\Delta t_p}}{2} = \frac{v}{2}\hat{t}_1 - \frac{v}{2}(t_0+T_s)
\end{equation}
where $v=1500$ m/sec is the (constant) speed of the acoustic waves underwater. Therefore, $z=\hat{d} \sim \mathcal{N}(d,\sigma_d^2)$, where $\sigma_d^2 = \frac{\sigma^2 v^2 PL}{ 16 P_T \mathbf{\dot{s}}^T \check{\mathbf{C}}^{-1} \mathbf{\dot{s}} } $. Let SNR$=1/\sigma^2$. Then, 
\begin{equation}
\label{eq:sigmad}
\sigma_d^2 = \frac{v^2 PL}{ 16 \text{SNR} P_T \mathbf{\dot{s}}^T \check{\mathbf{C}}^{-1} \mathbf{\dot{s}} }
\end{equation}

{\bf Remark 4.} The RTT-based distance estimation under the distance bounding protocol is commonly known as two-way ranging-based localization in the literature. We note that the two-way ranging-based localization schemes are (time) synchronization-free \cite{Mouftah:CST:2011},\cite{Poor:TWC:2010},\cite{Tan:JOE:2011}. In other words, RTT estimation only requires two timestamps $t_0$ and $t_1$ generated by the local oscillator/clock of the sink node; therefore, no explicit time synchronization among the UWASN nodes is needed. Nevertheless, we emphasize that the periodic broadcast of the challenge message by the sink node implicitly enables (coarse) time synchronization in the network. This is because the UWASN nodes then follow a master-slave architecture where the sink node acts as the master node, while the sensor nodes act as the slave nodes.

\subsection{Performance of Distance based Impersonation Detection and Transmitter Identification}
Test 1 (distance bounding test) first. Let $\sigma_{d_i}^2 = \frac{v^2 PL_i}{ 16 \text{SNR} P_T \mathbf{\dot{s}}^T \check{\mathbf{C}}^{-1} \mathbf{\dot{s}} }$ where $PL_i$ is the distance-dependent pathloss incurred by the transmission by $A_i$. Also, let $\sigma_{d_E}^2 = \frac{v^2 PL_E}{ 16 \text{SNR} P_T \mathbf{\dot{s}}^T \check{\mathbf{C}}^{-1} \mathbf{\dot{s}} }$ where $PL_E$ is the distance-dependent pathloss incurred by the transmission by Eve. Then, the probability of false alarm $P_{fa}$ is obtained by replacing $\sigma_d$ in Eq. (\ref{eq:pfa1}) with $\sigma_{d_i}$, while the probability of missed detection $P_{md}$ is obtained by replacing $\sigma_d$ in Eq. (\ref{eq:pmd1}) with $\sigma_{d_E}$. Note that $\sigma_{d_E}$ is a one-on-one function of $d_E$; therefore, everything else in Eq. (\ref{eq:pmd1}) remains intact.
Next, the Test 2(b). The $P_{fa}$ is obtained by replacing $\sigma_d$ with $\sigma_{d_i}$ in Eq. (\ref{eq:pfa2}). The $P_{md}$ is obtained by replacing $\sigma_d$ with $\sigma_{d_E}$ in Eq. (\ref{eq:pmd2}).
Finally, the transmitter identification. The probability of misclassification $P_{mc}$ is obtained by replacing $\sigma_d$ with $\sigma_{d_i}$ in Eq. (\ref{eq:pmc}).

\section{Performance Evaluation}
In this section, we first describe our simulation setup, and then present the simulation results which quantify the performance of the proposed impersonation detection framework for the AWGN UWA channel, and the UWA channel with colored noise and frequency-dependent pathloss.

\subsection{Simulation Setup} 
\begin{figure}
        \centering
        \begin{subfigure}[b]{0.5\textwidth}
                \includegraphics[width=\textwidth]{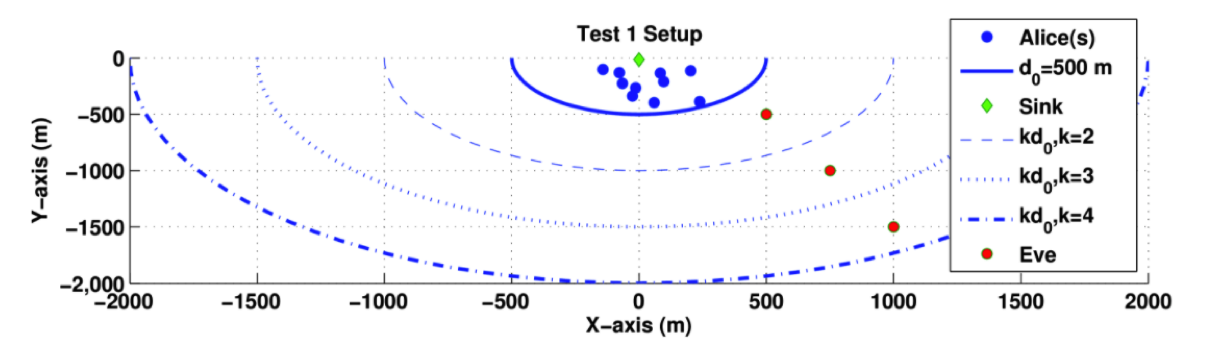}
                \caption{For \textit{step 1}, the Eve node is randomly placed outside the trusted zone, but within one of three half-discs (of radius $kd_0$ where $k>1$), one by one.}
                \label{fig:net}
        \end{subfigure}
        \begin{subfigure}[b]{0.5\textwidth}
                \includegraphics[width=\textwidth]{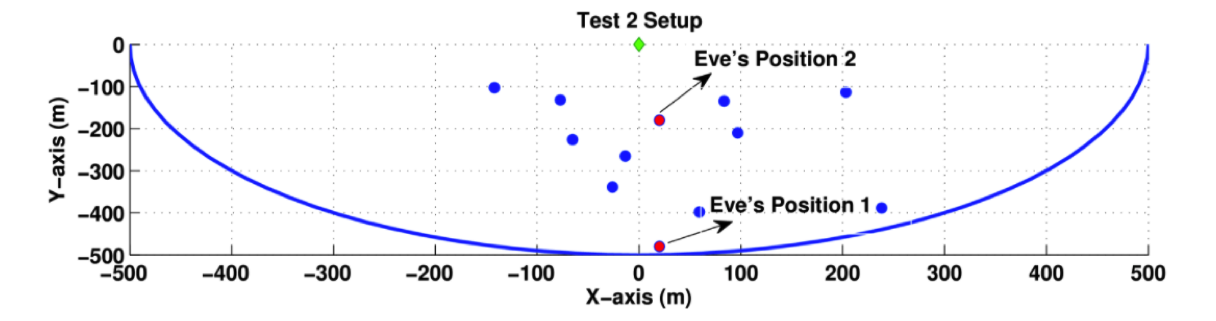}
                \caption{For \textit{step 2}, Eve is randomly placed at two different locations inside the trusted zone (where one location is close to, and other location is away from, the boundary of the trusted zone).}
                \label{fig:paths}
        \end{subfigure}
        \begin{subfigure}[b]{0.5\textwidth}
                \includegraphics[width=\textwidth]{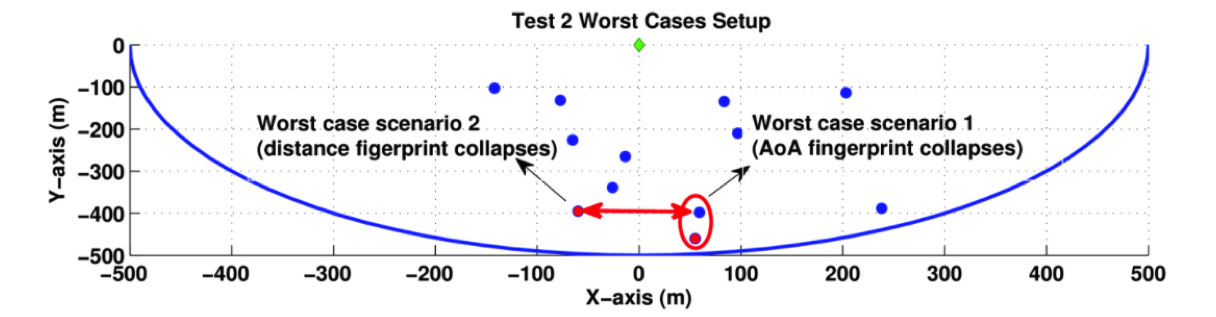}
                \caption{To demonstrate the strength of position based test (the\textit{ test 2(c)}), two worst case scenarios are considered where Eve node is strategically placed at two locations which culminate in either distance, or, AoA ceasing to be effective as the fingerprint of the sender node.}
                \label{fig:pu}
        \end{subfigure}
        \caption{Simulation Setup}
\label{fig:sim-setup}
\end{figure}

The performance evaluation was done in MATLAB. We consider a UWASN whose sensor (Alice) nodes are deployed in deep waters; and therefore, the reporting UWA channel is near-vertical (and thus, multipath-free). Fig. \ref{fig:sim-setup} shows the details of our simulation setup. The sink node is placed on the water surface at (0,0), while a trusted zone, in the shape of a vertical half-disc, of radius $d_0=500$ m is constructed around it ($d_0$ is set to $500$ m to realize a UWASN in deep waters). $M=10$ Alice nodes are deployed according to uniform distribution within the trusted zone. One Eve node is present which is randomly placed either outside the trusted zone, or, inside it (see Fig. \ref{fig:sim-setup}). The SNR at the sink node is defined as $1/\sigma^2$. For the AWGN UWA channel, we further assume that $\sigma_d^2=\sigma_{\theta}^2=\sigma_p^2=\sigma^2$ is the common estimation error corrupting the measurements of distance, AoA and position at the sink node\footnote{For simplicity of exposition, we assume that a mechanism to measure all the three features/fingerprints (i.e., distance, AoA and position) with the same quality exists. Furthermore, SNR as defined here does not represent quality of the underlying underwater UWA reporting channel; it rather is an indicator of the quality of a measurement.}. Such simplistic definition of SNR allows us to compare the performance of the various hypothesis tests and fusion rules proposed in Algorithm 1 against each other. On the other hand, for the UWA channel with colored noise (with covariance matrix $\mathbf{C}=\sigma^2 \check{\mathbf{C}}$) and frequency-dependent pathloss $PL(d,f)$, $ \sigma_d^2 =\bigg( \frac{v^2 PL(d,f)}{ 16 P_T \mathbf{\dot{s}}^T \check{\mathbf{C}}^{-1} \mathbf{\dot{s}} }\bigg)\sigma^2 $.


\subsection{Simulation Results: AWGN UWA Channel} 

\begin{figure}
\begin{center}
\includegraphics[width=10cm, height=7cm]{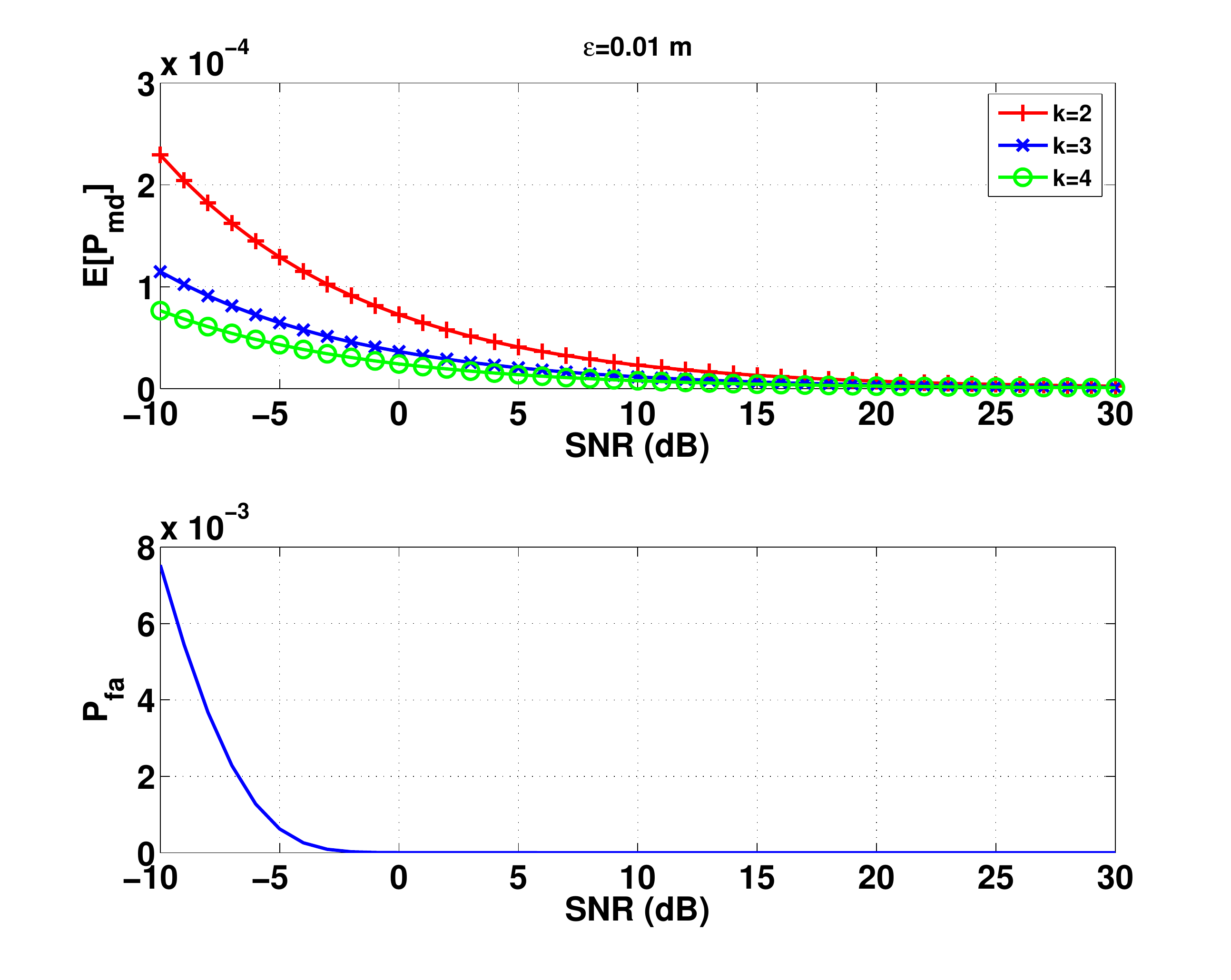}
\caption{Impersonation detection performance of \textit{step 1}: Eve is placed outside the trusted zone (as depicted in Fig. \ref{fig:sim-setup} (a)). {\it Both classification errors approach zero exponentially as the SNR is increased}.}
\label{fig:test1-pfapmdvssnr}
\end{center}
\end{figure}

\begin{figure}
\includegraphics[width=8.5cm]{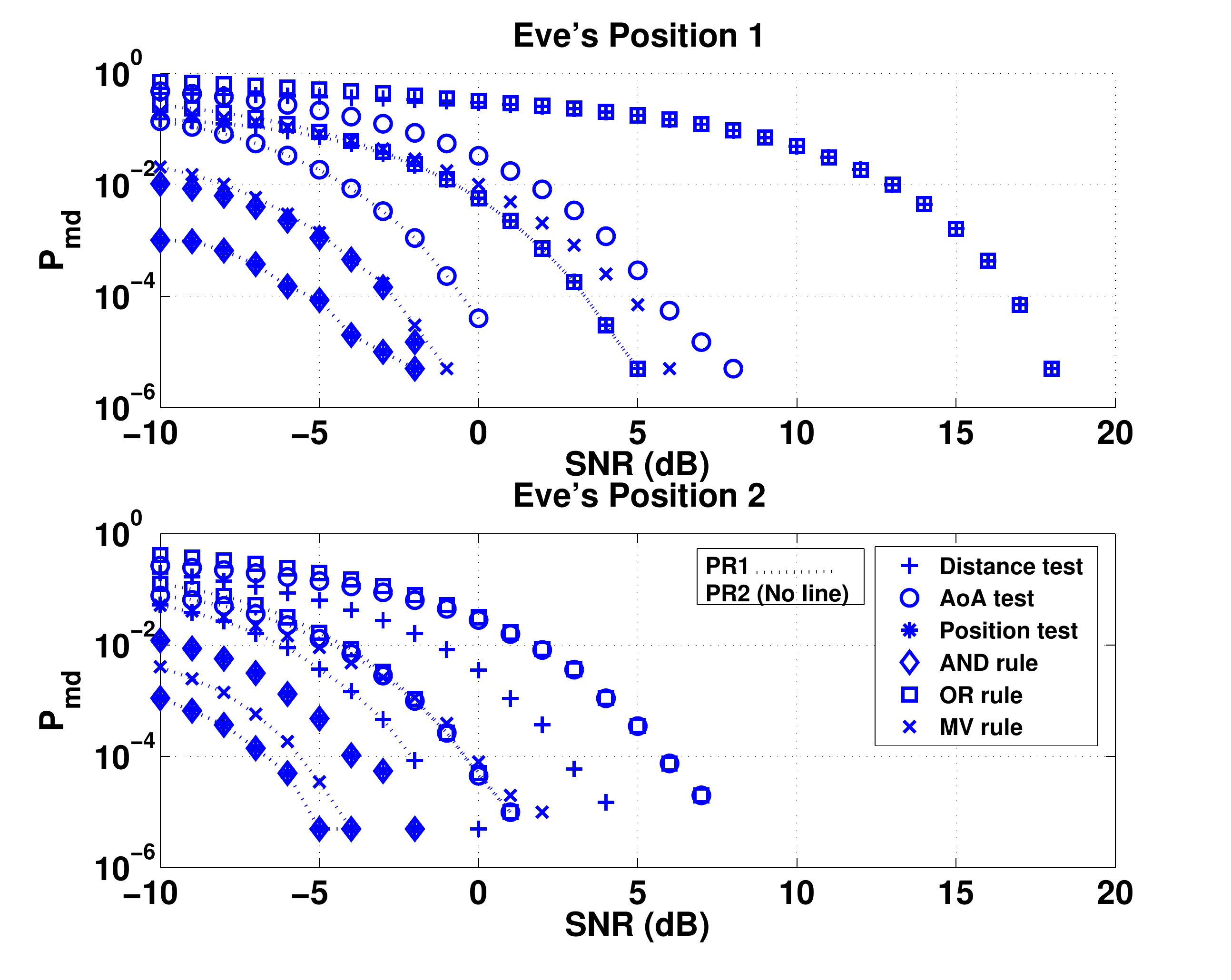}
\caption{Impersonation detection performance of \textit{step 2}: Eve is placed at two locations inside the trusted zone (as depicted in Fig. \ref{fig:sim-setup} (b)); PR stands for proximity region; for PR1, $\epsilon_p=1m^2$, $\epsilon_d=1$m, $\epsilon_\theta=1^{\circ}$; for PR2, $\epsilon_p=3m^2$, $\epsilon_d=3$m, $\epsilon_\theta=3^{\circ}$. {\it The success probability of Eve vanishes as the SNR is increased}. }
\label{fig:test2-pmdvssnr}
\end{figure}

\begin{figure}
\includegraphics[width=8.5cm]{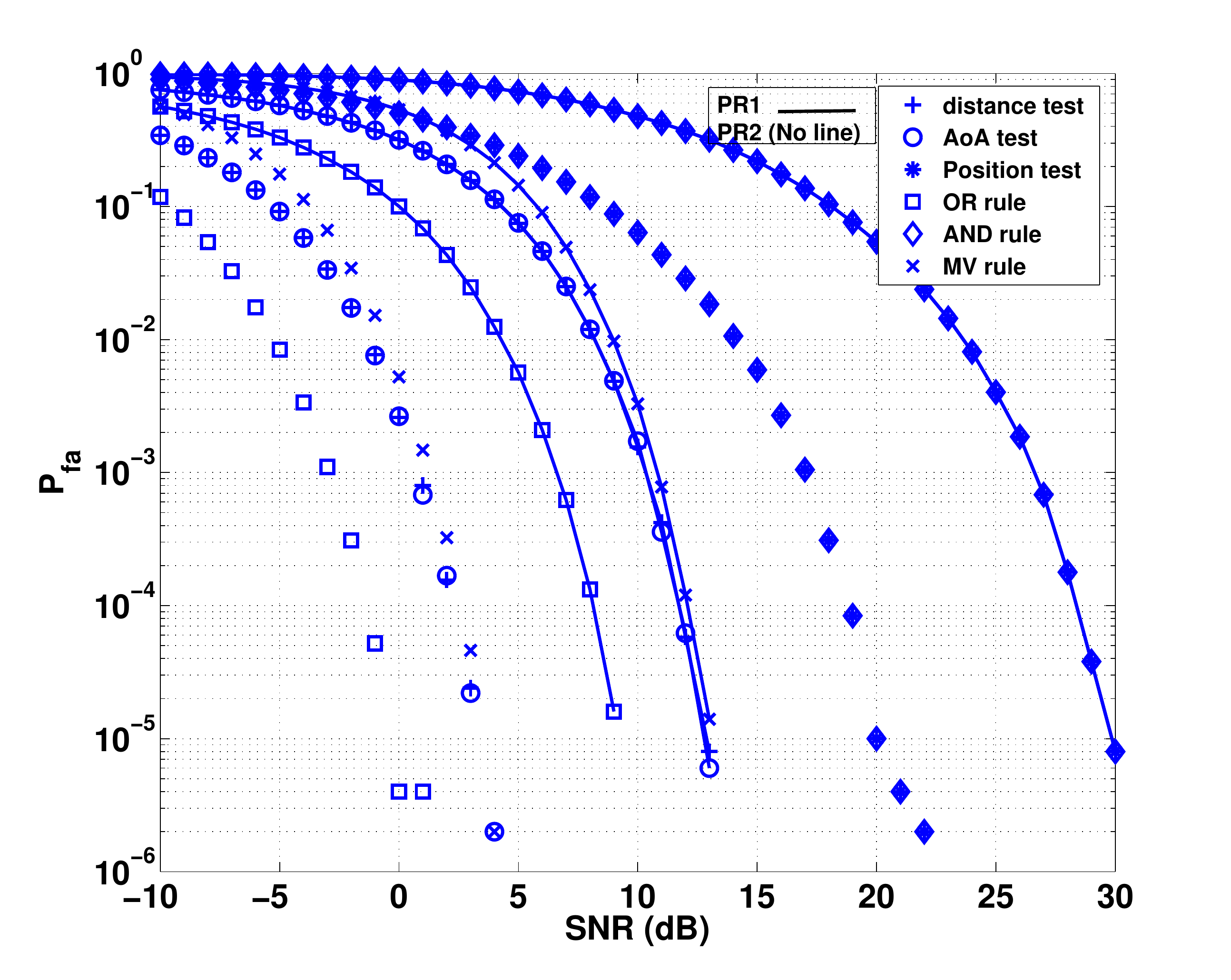}
\caption{Impersonation detection performance of \textit{step 2}: Eve is placed inside the trusted zone; for PR1, $\epsilon_p=1m^2$, $\epsilon_d=1$m, $\epsilon_\theta=1^{\circ}$; for PR2, $\epsilon_p=3m^2$, $\epsilon_d=3$m, $\epsilon_\theta=3^{\circ}$. {\it The false alarm rate vanishes to zero with an increase in the SNR}. }
\label{fig:test2-pfavssnr}
\end{figure}

\begin{figure}
\includegraphics[width=9cm]{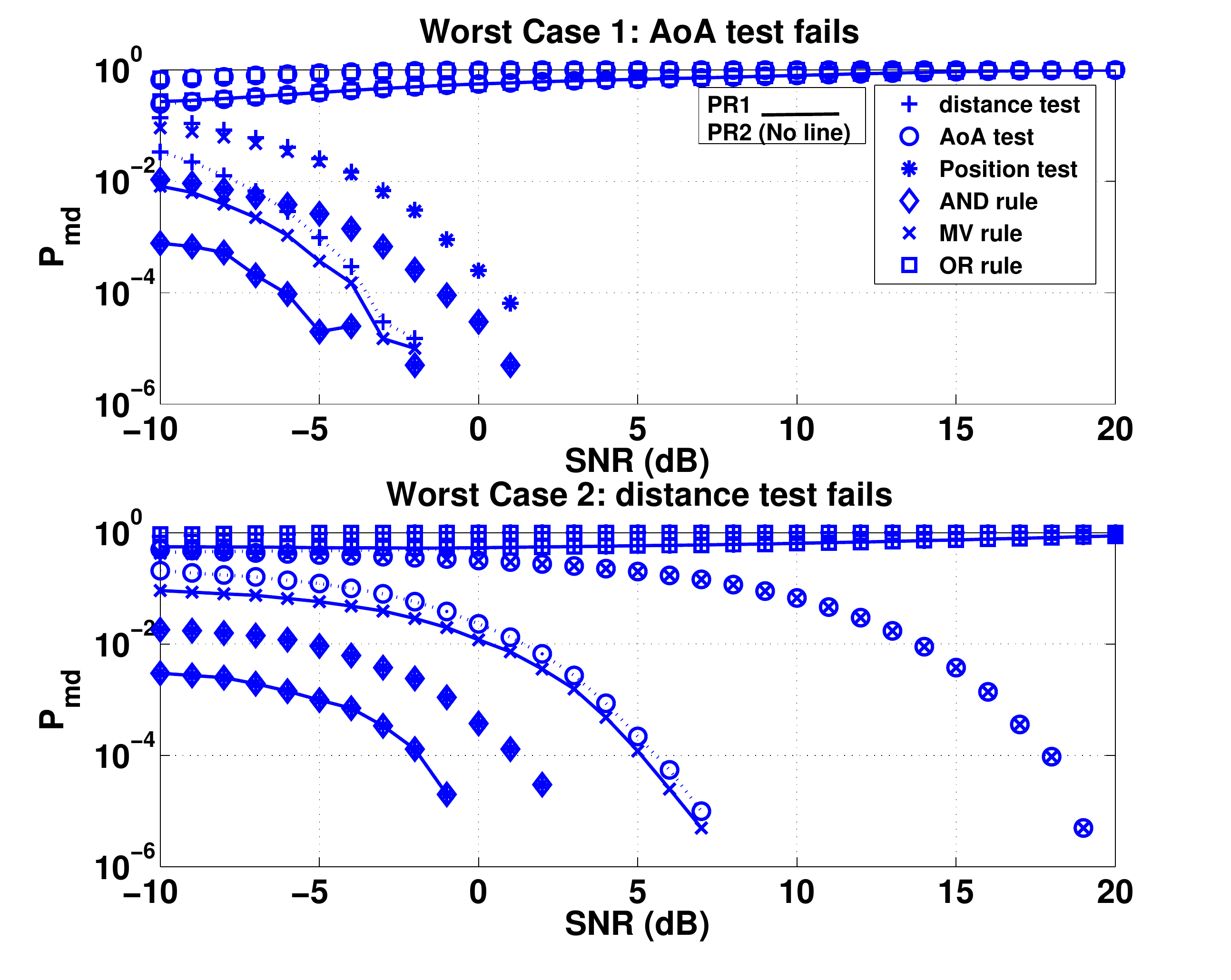}
\caption{Worst-case impersonation detection performance of \textit{step 2}: Eve is strategically placed at two locations inside the trusted zone (as depicted in Fig. \ref{fig:sim-setup} (c)); for PR1, $\epsilon_p=1$ $m^2$, $\epsilon_d=1$ m, $\epsilon_\theta=1^{\circ}$; for PR2, $\epsilon_p=3$ $m^2$, $\epsilon_d=3$ m, $\epsilon_\theta=3^{\circ}$. {\it For each of the two positions of Eve, either distance or AoA ceases to be effective as device fingerprint as its missed detection rate approaches one at high SNR values; however, location remains effective as fingerprint as its missed detection rate approaches zero at high SNRs}. }
\label{fig:test2-worstcase}
\end{figure}

\begin{figure}
\includegraphics[width=9cm]{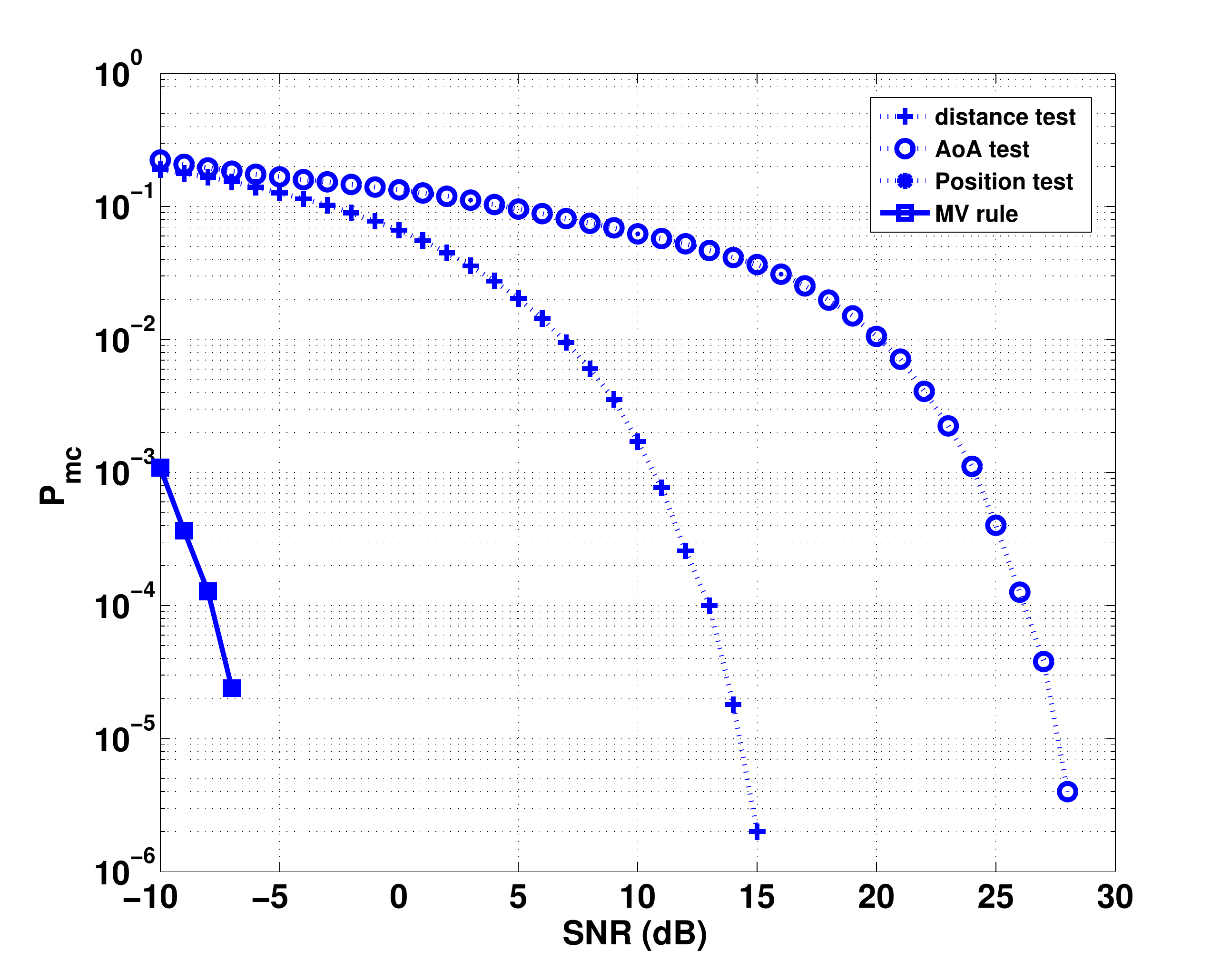}
\caption{Transmitter identification performance of \textit{step 2}: {\it The misclassification rate reduces to zero with increase in SNR}. (The curve for the MV rule is superimposed on the curve for the position test).}
\label{fig:test2-txidvssnr}
\end{figure}

Fig. \ref{fig:test1-pfapmdvssnr} investigates the impersonation detection performance of step 1, Figs. \ref{fig:test2-pmdvssnr}, \ref{fig:test2-pfavssnr}, \ref{fig:test2-worstcase} together investigate the impersonation detection performance of step 2, while Fig. \ref{fig:test2-txidvssnr} investigates the transmitter identification performance of step 2, for the AWGN UWA channel. 

Fig. \ref{fig:test1-pfapmdvssnr} plots the impersonation detection performance of step 1 (the distance bounding test). To obtain the results in Fig. \ref{fig:test1-pfapmdvssnr}, Eve is randomly placed at three different locations outside the trusted zone (see Fig. \ref{fig:sim-setup} (a)). Specifically, Fig. \ref{fig:test1-pfapmdvssnr} sketches the tradeoff of the two error probabilities ($P_{md}$, $P_{fa}$) against the SNR whereby both $P_{md}$ \& $P_{fa}$ decrease with an increase in SNR. However, since the centroid of the Alice nodes' positions (for the deployment shown in Fig. \ref{fig:sim-setup}) is away from the boundary of the trusted zone, $P_{fa}$ vanishes (to zero) much faster with an increase in SNR.

Fig. \ref{fig:test2-pmdvssnr} studies the decay rate of the success probability of Eve ($P_{md}$) as a function of SNR. To obtain the results in Fig. \ref{fig:test2-pmdvssnr}, Eve is randomly placed at two different locations within the trusted zone (see Fig. \ref{fig:sim-setup} (b)). As anticipated, the AND (OR) rule being a pessimistic (optimistic) rule performs the best (worst). More precisely, for any given SNR, the AND (OR) rule minimizes (maximizes) the $P_{md}$; equivalently, for any given requirement on $P_{md}$, the AND (OR) rule requires much lesser (higher) SNR compared to the other schemes. Additionally, the performance of the Position test is identical to that of AND rule (this is because the position/location, by definition, is the AND/combining of distance and AoA). Lastly, increasing the area of the proximity region for each of the tests 2(a), 2(b), 2(c) results in degradation of the detection performance of step 2.

Fig. \ref{fig:test2-pfavssnr} plots the probability of false alarm $P_{fa}$ (an indicator of data rate shrinkage)\footnote{False alarm, by definition, is the case when the sink node ends up discarding the data from the legitimate (Alice) nodes, which results in reduction in net data rate, increased latency due to re-transmissions, etc.} as a function of SNR. Once again, the OR (AND) rule performs the best (worst) as anticipated. This is because the OR (AND) rule, by definition, minimizes (maximizes) the probability of false alarm. Furthermore, the performance of the Position test (test 2(a)) coincides with the performance of the AND rule. Finally, increasing the area of the proximity region for each of the tests 2(a), 2(b), 2(c) results in reduction in the probability of data rate shrinkage, as expected.

Fig. \ref{fig:test2-worstcase} captures the so-called worst case scenarios for test 2 whereby some individual (specifically, the weaker one) fingerprints collapse. Specifically, the first worst case scenario considers the situation where $|\theta_E-\theta_i|<\epsilon_\theta$ indefinitely (see Fig. \ref{fig:sim-setup} (c)). Therefore, in this situation, AoA ceases to be effective as the fingerprint of the transmit device. In such situation, SNR becomes a foe instead of a friend, i.e., $\underset{{\mathrm{SNR}\rightarrow \infty}}{\lim} P_{md}^{(AoA)}=1$ (see the top plot of Fig. \ref{fig:test2-worstcase}). 
Similarly, the second worst case scenario captures the situation where $|d_E-d_i|<\epsilon_d$ indefinitely (see Fig. \ref{fig:sim-setup} (c)) which culminates in distance being ineffective as fingerprint of the transmit device. Once again, an increase in SNR makes the situation worse, i.e., $\underset{{\mathrm{SNR}\rightarrow \infty}}{\lim} P_{md}^{(d)}=1$ (see the bottom plot in Fig. \ref{fig:test2-worstcase}). However, one can see that the Position test as well as AND rule gracefully sustain such worst case scenarios\footnote{The scenario $|p_E-p_i|<\epsilon_p$ is omitted simply because it implies that the Eve is co-located with some $A_i$ (assuming that $\epsilon_p$ equals the size of a typical UWASN node).}.

Fig. \ref{fig:test2-txidvssnr} plots the decay rate of the misclassification error $P_{mc}$ (i.e., incorrectly identifying Alice $i$ as Alice $j$) against SNR for all the three tests 2(a), 2(b), 2(c), and their fusion via MV rule. From Fig. \ref{fig:test2-txidvssnr}, one can see that the Position test outperforms the other two tests (distance based, AoA based) by a big margin, while the curve for the MV rule is superimposed on the curve for the Position test. This is expected, because as explained in Remark 3, the proximity region of the Position test is much smaller than the proximity regions of the distance test and the AoA test.

\subsection{Simulation Results: UWA Channel with Colored Noise and Frequency-dependent Pathloss} 

\begin{figure}
\begin{center}
\includegraphics[width=10cm, height=7cm]{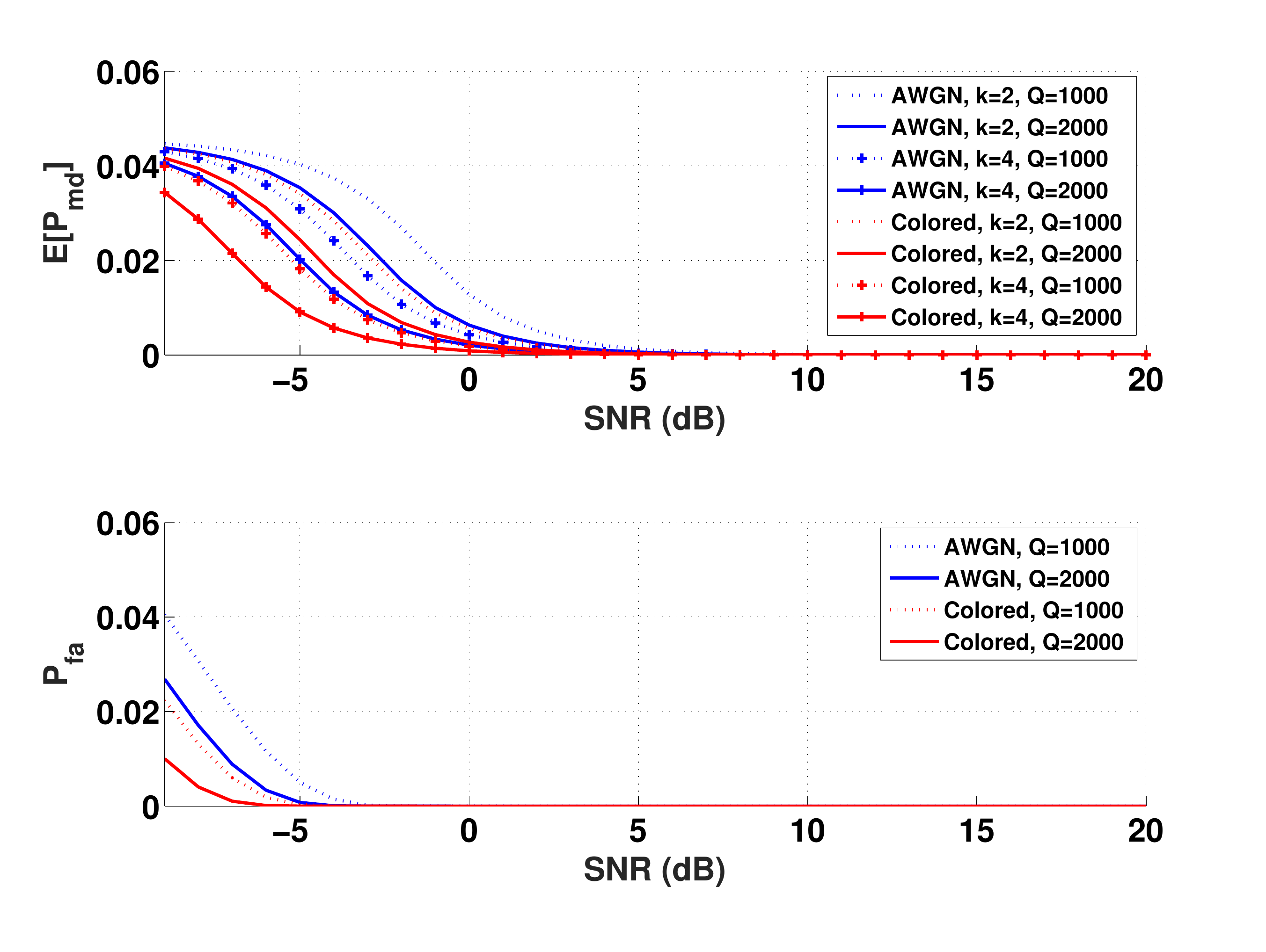}
\caption{Impersonation detection performance of \textit{step 1}: Eve is placed outside the trusted zone (as depicted in Fig. \ref{fig:sim-setup} (a)). {\it Both error probabilities reduce to zero with increase in SNR. Moreover, the performance of step 1 increases as we collect more samples $Q$ for distance estimation. Finally, the performance of step 1 is better in UWA channel with colored noise than the AWGN UWA channel.} }
\label{fig:test1-pfapmdvssnrcolored}
\end{center}
\end{figure}

\begin{figure}
\begin{center}
\includegraphics[width=10cm, height=7cm]{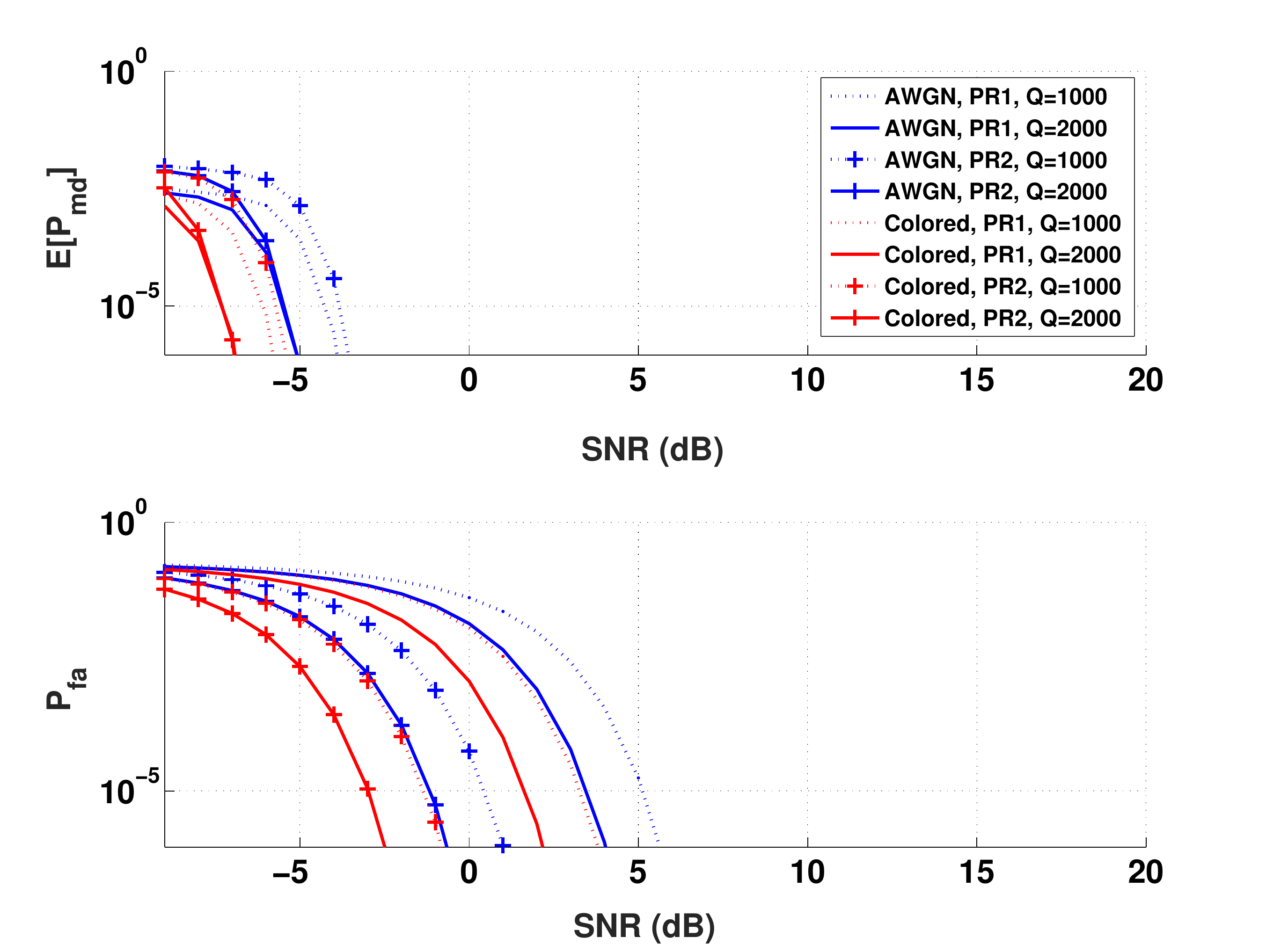}
\caption{Impersonation detection performance of \textit{step 2}: Eve is placed inside the trusted zone; for PR1, $\epsilon_d=1$m; for PR2, $\epsilon_d=3$m. {\it Both error probabilities reduce to zero with increase in SNR. Additionally, increasing the area of the proximity region results in degradation in detection performance of step 2. Finally, the threshold SNR to achieve arbitrarily small errors for $P_{md}$ is about 5 dB lower than that for $P_{fa}$.} }
\label{fig:test2-pfapmdvssnrcolored}
\end{center}
\end{figure}

\begin{figure}
\begin{center}
\includegraphics[width=10cm, height=7cm]{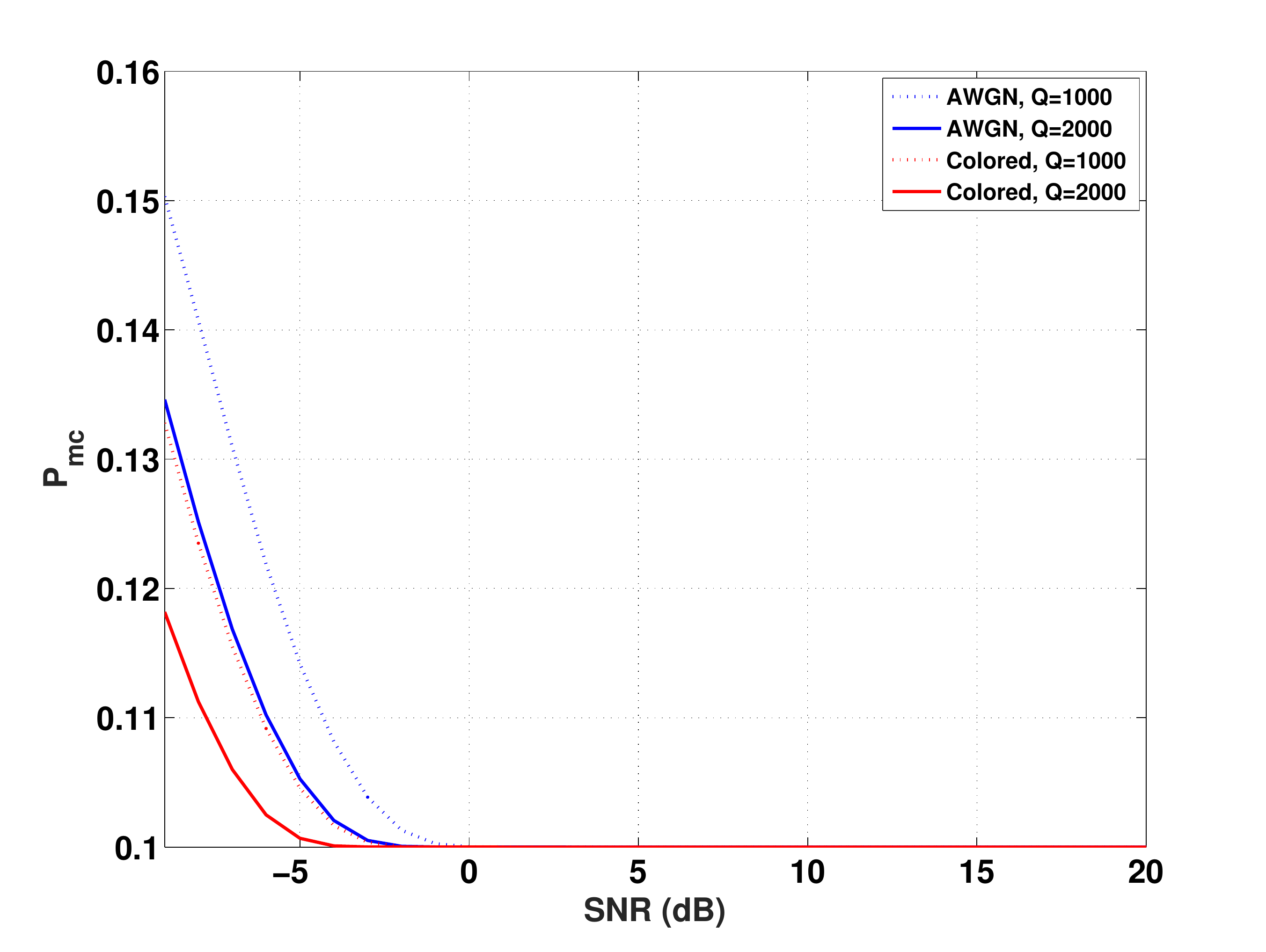}
\caption{Transmitter identification performance of \textit{step 2}. {\it The misclassification rate reduces to zero with increase in SNR. Moreover, the performance increases as we collect more samples $Q$ for distance estimation. Finally, the performance is better in UWA channel with colored noise than the AWGN UWA channel. } }
\label{fig:misclasscolored}
\end{center}
\end{figure}

Figs. \ref{fig:test1-pfapmdvssnrcolored}, \ref{fig:test2-pfapmdvssnrcolored}, \ref{fig:misclasscolored} investigate the impersonation detection performance of step 1, step 2, and transmitter identification performance of step 2 respectively, for the UWA Channel with colored noise and frequency-dependent pathloss. For the sake of fair comparison, we set $\check{\mathbf{C}}=\mathbf{I}$, or, $\mathbf{C}=\sigma^2\mathbf{I}$ to realize an AWGN UWA channel that is exposed to frequency-dependent pathloss ($\mathbf{I}$ is a $Q\times Q$ identity matrix). This AWGN UWA channel, therefore, is different than the AWGN UWA channel considered in Section V-B which sees no pathloss. For all the plots in this sub-section, we set $P_T=250$ dB$\mu$ Pascals.

Fig. \ref{fig:test1-pfapmdvssnrcolored} plots the two error probabilities of step 1 (the distance bounding test) against SNR. To our surprise, the proposed impersonation detection scheme performs better in UWA channel with colored noise than in AWGN UWA channel. Additionally, as expected, an increase in $Q$ (i.e., collecting more samples for estimation during a slot) culminates in better performance (due to better distance estimate). Last but not the least, for both channels (AWGN UWA, and UWA with colored noise), the threshold SNR to achieve arbitrarily small errors for the probability of false alarm is about $5$ dB lower than the probability of missed detection.

Fig. \ref{fig:test2-pfapmdvssnrcolored} studies the impersonation detection performance of step 2 (test 2(b)) as a function of SNR. We once again notice that an increase in $Q$ (reduces both kind of errors at any given SNR, and thus) leads to improved performance, and that the proposed test 2(b) performs better in the face of colored noise. We also note that increasing the area of the proximity region for the test 2(b) results in degradation of the detection performance of test 2(b). Finally, for step 2, the probability of missed detection drops to zero much faster than the probability of false alarm. That is, the threshold SNR to achieve arbitrarily small errors for $P_{md}$ is at least $5$ dB lower than $P_{fa}$.

Fig. \ref{fig:misclasscolored} plots the decay rate of the misclassification error $P_{mc}$ (i.e., incorrectly identifying $A_i$ as $A_j$) against SNR for test 2(b). This result corroborates our earlier observations in Figs. \ref{fig:test1-pfapmdvssnrcolored}, \ref{fig:test2-pfapmdvssnrcolored}, i.e., an increase in $Q$ leads to improved performance, and that the proposed test 2(b) performs better in the face of colored noise.

\subsection{Discussions}

\begin{itemize}

\item
The results in Figs. \ref{fig:test2-pmdvssnr}, \ref{fig:test2-pfavssnr}, \ref{fig:test2-worstcase} indicate that, under the impersonation detection problem, it is not possible to minimize both $P_{md}$ and $P_{fa}$ at the same time because of their conflicting nature. In other words, one could minimize one error type only by compromising on the other error type (which is inline with Neyman-Pearson Theorem \cite{Yan:TIT:2001}). 

\item
To our surprise, Figs. \ref{fig:test1-pfapmdvssnrcolored}, \ref{fig:test2-pfapmdvssnrcolored}, \ref{fig:misclasscolored} reveal that the proposed impersonation detection (and transmitter identification) scheme performs better in UWA channel with colored noise than in the AWGN UWA channel. Looking at Eq. (\ref{eq:sigmad}), we see that only $\check{\mathbf{C}}$ changes in going from an AWGN UWA channel to a UWA channel with colored noise (while all the other parameters stay the same); therefore, the behavior observed is mainly due to change in the Frobenius norm of $\check{\mathbf{C}}^{-1}$. In a nutshell, this finding prompts us to the optimistic conclusion that the proposed method is indeed effective for a UWA channel with colored noise and frequency-dependent pathloss.

\item
This work does not have experimental results to report to support the simulation results presented earlier. Nevertheless, the reader interested in experimental validation of the proposed impersonation detection framework is referred to the works \cite{Parra:SJ:2016,Song:JOE:2011,Song:makai:2006}. Specifically, \cite{Parra:SJ:2016} summarizes the state-of-the-art in commercial underwater acoustic modems, while the details pertinent to the (commercially available) arrays of hydrophones could be found in \cite{Song:JOE:2011},\cite{Song:makai:2006} which report experimental results. 

\item
Though this work assumes a 2D geometry/deployment of the UWASN nodes for the sake of clarity of exposition, extension of the proposed impersonation detection framework to the case of 3D geometry/deployment of the UWASN nodes is laborious but straightforward. Yet, some comments are in order. Under the 3D geometry, the sink node will have to estimate two angles of arrival, the azimuth AoA $\theta$ and elevation AoA $\phi$ in addition to the distance estimation. For this purpose, the sink node could utilize a uniform circular array instead of a uniform linear array. With the distance estimate and the estimates of the two AoAs available, the sink node could then uniquely estimate the location/position of the transmit node as $(d,\theta,\phi)$ in spherical coordinates. Furthermore, the additional angle of arrival could serve as an additional feature. But the overall framework (as summarized in Fig. 4 and Algorithm 1) remains the same as before.   

\end{itemize}

\section{Conclusion \& Future Work} \label{sec:con}

This work addressed the problem of impersonation attack detection in a line-of-sight underwater acoustic sensor network (UWASN), for both additive white Gaussian noise (AWGN)-limited underwater acoustic (UWA) channel, and the UWA channel with colored noise and frequency-dependent pathloss. We first proposed a novel, multiple-features based, two-step method which utilized the distance, the angle of arrival (AoA), and the location of a sender node as device fingerprints to carry out the authentication as well as the transmitter identification, for the AWGN UWA channel. To this end, we provided closed-form expressions for the error probabilities (i.e., the performance) of most of the hypothesis tests. We then considered the case of a UWA with colored noise and frequency-dependent pathloss, and derived a maximum-likelihood distance estimator as well as the corresponding Cramer-Rao bound. We then invoked the proposed two-step, impersonation detection framework by utilizing distance as the sole feature. Simulation results verified the feasibility of the proposed scheme when applied to a UWA channel with colored noise and frequency-dependent pathloss.

This work opens up many interesting possibilities for future work. For example, when the Eve and/or Alice nodes are mobile, a Bayesian filtering framework (such as \cite{Mahboob:Globecom:2014}) could be employed to track the motion of each mobile sensor node to keep up with the need of obtaining the updated ground truth periodically. Additionally, a more general scenario whereby multiple Eve nodes (with the exact count of Eve nodes not known a priori) are present need to be studied. Finally, adapting the proposed method to more complex scenarios, e.g., multipath propagation, reverberation, stratification etc., is yet another promising direction of research.
 
\appendices

\footnotesize{
\bibliographystyle{IEEEtran}
\bibliography{references}
}

\vfill\break

\end{document}